\magnification=\magstep1
\advance\vsize by 1pc
\input psfig.sty


\catcode`\@=11


\message{Loading a modification of the jyTeX macros...}

\message{modifications to plain.tex,}


\def\newcount{\alloc@0\count\countdef\insc@unt}
\def\newdimen{\alloc@1\dimen\dimendef\insc@unt}
\def\newskip{\alloc@2\skip\skipdef\insc@unt}
\def\newmuskip{\alloc@3\muskip\muskipdef\@cclvi}
\def\newtoks{\alloc@5\toks\toksdef\@cclvi}
\def\newhelp#1#2{\newtoks#1\global#1\expandafter{\csname#2\endcsname}}
\def\newread{\alloc@6\read\chardef\sixt@@n}
\def\newwrite{\alloc@7\write\chardef\sixt@@n}
\def\newfam{\alloc@8\fam\chardef\sixt@@n}
\def\newinsert#1{\global\advance\insc@unt by\m@ne
     \ch@ck0\insc@unt\count
     \ch@ck1\insc@unt\dimen
     \ch@ck2\insc@unt\skip
     \ch@ck4\insc@unt\box
     \allocationnumber=\insc@unt
     \global\chardef#1=\allocationnumber
     \wlog{\string#1=\string\insert\the\allocationnumber}}
\def\newif#1{\count@\escapechar \escapechar\m@ne
     \expandafter\expandafter\expandafter
          \xdef\@if#1{true}{\let\noexpand#1=\noexpand\iftrue}%
     \expandafter\expandafter\expandafter
          \xdef\@if#1{false}{\let\noexpand#1=\noexpand\iffalse}%
     \global\@if#1{false}\escapechar=\count@}


\newlinechar=`\^^J
\overfullrule=0pt

\message{hacks,}


\toksdef\toks@i=1
\toksdef\toks@ii=2


\def\TeX{T\kern-.1667em \lower.5ex \hbox{E}\kern-.125em X\null}
\def\jyTeX{{\leavevmode
     \raise.587ex \hbox{\it\j}\kern-.1em \lower.048ex \hbox{\it y}\kern-.12em
     \TeX}}

\let\then=\iftrue
\def\ifnoarg#1\then{\def\hack@{#1}\ifx\hack@\empty}
\def\ifundefined#1\then{%
     \expandafter\ifx\csname\expandafter\blank\string#1\endcsname\relax}
\def\useif#1\then{\csname#1\endcsname}
\def\usename#1{\csname#1\endcsname}
\def\useafter#1#2{\expandafter#1\csname#2\endcsname}

\long\def\loop#1\repeat{\def\@iterate{#1\expandafter\@iterate\fi}\@iterate
     \let\@iterate=\relax}

\let\TeXend=\end
\def\begin#1{\begingroup\def\@@blockname{#1}\usename{begin#1}}
\def\End#1{\usename{end#1}\def\hack@{#1}%
     \ifx\@@blockname\hack@
          \endgroup
     \else\err@badgroup\hack@\@@blockname
     \fi}
\def\@@blockname{}

\def\defaultoption[#1]#2{%
     \def\hack@{\ifx\hack@ii[\toks@={#2}\else\toks@={#2[#1]}\fi\the\toks@}%
     \futurelet\hack@ii\hack@}

\def\markup#1{\let\@@marksf=\empty
     \ifhmode\edef\@@marksf{\spacefactor=\the\spacefactor\relax}\/\fi
     ${}^{\hbox{\subscriptfonts#1}}$\@@marksf}


\newtoks\shortyear
\newtoks\militaryhour
\newtoks\standardhour
\newtoks\minute
\newtoks\amorpm

\def\settime{\count@=\time\divide\count@ by60
     \militaryhour=\expandafter{\number\count@}%
     {\multiply\count@ by-60 \advance\count@ by\time
          \xdef\hack@{\ifnum\count@<10 0\fi\number\count@}}%
     \minute=\expandafter{\hack@}%
     \ifnum\count@<12
          \amorpm={am}
     \else\amorpm={pm}
          \ifnum\count@>12 \advance\count@ by-12 \fi
     \fi
     \standardhour=\expandafter{\number\count@}%
     \def\hack@19##1##2{\shortyear={##1##2}}%
          \expandafter\hack@\the\year}

\def\monthword#1{%
     \ifcase#1
          $\bullet$\err@badcountervalue{monthword}%
          \or January\or February\or March\or April\or May\or June%
          \or July\or August\or September\or October\or November\or December%
     \else$\bullet$\err@badcountervalue{monthword}%
     \fi}

\def\monthabbr#1{%
     \ifcase#1
          $\bullet$\err@badcountervalue{monthabbr}%
          \or Jan\or Feb\or Mar\or Apr\or May\or Jun%
          \or Jul\or Aug\or Sep\or Oct\or Nov\or Dec%
     \else$\bullet$\err@badcountervalue{monthabbr}%
     \fi}

\def\militarytime{\the\militaryhour:\the\minute}
\def\standardtime{\the\standardhour:\the\minute}


\def\@setnumstyle#1#2{\expandafter\global\expandafter\expandafter
     \expandafter\let\expandafter\expandafter
     \csname @\expandafter\blank\string#1style\endcsname
     \csname#2\endcsname}
\def\numstyle#1{\usename{@\expandafter\blank\string#1style}#1}
\def\ifblank#1\then{\useafter\ifx{@\expandafter\blank\string#1}\blank}

\def\blank#1{}

\def\Roman#1{\expandafter\uppercase\expandafter{\romannumeral#1}}
\def\alphabetic#1{%
     \ifcase#1
          $\bullet$\err@badcountervalue{alphabetic}%
          \or a\or b\or c\or d\or e\or f\or g\or h\or i\or j\or k\or l\or m%
          \or n\or o\or p\or q\or r\or s\or t\or u\or v\or w\or x\or y\or z%
     \else$\bullet$\err@badcountervalue{alphabetic}%
     \fi}
\def\Alphabetic#1{\expandafter\uppercase\expandafter{\alphabetic{#1}}}
\def\symbols#1{%
     \ifcase#1
          $\bullet$\err@badcountervalue{symbols}%
          \or*\or\dag\or\ddag\or\S\or$\|$%
          \or**\or\dag\dag\or\ddag\ddag\or\S\S\or$\|\|$%
     \else$\bullet$\err@badcountervalue{symbols}%
     \fi}


\catcode`\^^?=13 \def^^?{\relax}

\def\trimleading#1\to#2{\edef#2{#1}%
     \expandafter\@trimleading\expandafter#2#2^^?^^?}
\def\@trimleading#1#2#3^^?{\ifx#2^^?\def#1{}\else\def#1{#2#3}\fi}

\def\trimtrailing#1\to#2{\edef#2{#1}%
     \expandafter\@trimtrailing\expandafter#2#2^^? ^^?\relax}
\def\@trimtrailing#1#2 ^^?#3{\ifx#3\relax\toks@={}%
     \else\def#1{#2}\toks@={\trimtrailing#1\to#1}\fi
     \the\toks@}

\def\trim#1\to#2{\trimleading#1\to#2\trimtrailing#2\to#2}

\catcode`\^^?=15


\long\def\additemL#1\to#2{\toks@={\^^\{#1}}\toks@ii=\expandafter{#2}%
     \xdef#2{\the\toks@\the\toks@ii}}

\long\def\additemR#1\to#2{\toks@={\^^\{#1}}\toks@ii=\expandafter{#2}%
     \xdef#2{\the\toks@ii\the\toks@}}

\def\getitemL#1\to#2{\expandafter\@getitemL#1\hack@#1#2}
\def\@getitemL\^^\#1#2\hack@#3#4{\def#4{#1}\def#3{#2}}


\newskip\headskip
\newskip\footskip

\message{document layout,}

\newif\ifdraft
\def\draft{\drafttrue\leftmargin=.5in \overfullrule=5pt }


\newskip\abovechapterskip
\newskip\belowchapterskip
\newskip\abovesectionskip
\newskip\belowsectionskip
\newskip\abovesubsectionskip
\newskip\belowsubsectionskip

\def\chapterstyle#1{\global\expandafter\let\expandafter\@chapterstyle
     \csname#1text\endcsname}
\def\sectionstyle#1{\global\expandafter\let\expandafter\@sectionstyle
     \csname#1text\endcsname}
\def\subsectionstyle#1{\global\expandafter\let\expandafter\@subsectionstyle
     \csname#1text\endcsname}

\def\CHapter#1{%
     \ifdim\lastskip=17sp \else\chapterbreak\vskip\abovechapterskip\fi
     \@chapterstyle{\ifblank\chapternumstyle\then
          \else\newchapternum=\next\chapternumformat\ \fi#1}%
     \nobreak\vskip\belowchapterskip\vskip17sp }

\def\Section#1{%
     \ifdim\lastskip=17sp \else\sectionbreak\vskip\abovesectionskip\fi
     \@sectionstyle{\ifblank\sectionnumstyle\then
          \else\newsectionnum=\next\sectionnumformat\ \fi#1}%
     \nobreak\vskip\belowsectionskip\vskip17sp }

\def\Subsection#1{%
     \ifdim\lastskip=17sp \else\subsectionbreak\vskip\abovesubsectionskip\fi
     \@subsectionstyle{\ifblank\subsectionnumstyle\then
          \else\newsubsectionnum=\next\subsectionnumformat\ \fi#1}%
     \nobreak\vskip\belowsubsectionskip\vskip17sp }


\newtoks\everybye \everybye={\par\vfil}
\outer\def\bye{\the\everybye
     \footnotecheck
     \prelabelcheck
     \streamcheck
     \supereject
     \TeXend}

\message{labels,}

\let\@@labeldef=\xdef
\newif\if@labelfile
\newwrite\@labelfile
\let\@prelabellist=\empty

\def\Label#1#2{\trim#1\to\@@labarg\edef\@@labtext{#2}%
     \edef\@@labname{lab@\@@labarg}%
     \useafter\ifundefined\@@labname\then\else\@yeslab\fi
     \useafter\@@labeldef\@@labname{#2}%
     \ifstreaming
          \expandafter\toks@\expandafter\expandafter\expandafter
               {\csname\@@labname\endcsname}%
          \immediate\write\streamout{\noexpand\Label{\@@labarg}{\the\toks@}}%
     \fi}
\def\@yeslab{%
     \useafter\ifundefined{if\@@labname}\then
          \err@labelredef\@@labarg
     \else\useif{if\@@labname}\then
               \err@labelredef\@@labarg
          \else\global\usename{\@@labname true}%
               \useafter\ifundefined{pre\@@labname}\then
               \else\useafter\ifx{pre\@@labname}\@@labtext
                    \else\err@badlabelmatch\@@labarg
                    \fi
               \fi
               \if@labelfile
               \else\global\@labelfiletrue
                    \immediate\write\sixt@@n{--> Creating file \jobname.lab}%
                    \immediate\openout\@labelfile=\jobname.lab
               \fi
               \immediate\write\@labelfile
                    {\noexpand\prelabel{\@@labarg}{\@@labtext}}%
          \fi
     \fi}

\def\putlab#1{\trim#1\to\@@labarg\edef\@@labname{lab@\@@labarg}%
     \useafter\ifundefined\@@labname\then\@nolab\else\usename\@@labname\fi}
\def\@nolab{%
     \useafter\ifundefined{pre\@@labname}\then
          \undefinedlabelformat
          \err@needlabel\@@labarg
          \useafter\xdef\@@labname{\undefinedlabelformat}%
     \else\usename{pre\@@labname}%
          \useafter\xdef\@@labname{\usename{pre\@@labname}}%
     \fi
     \useafter\newif{if\@@labname}%
     \expandafter\additemR\@@labarg\to\@prelabellist}

\def\prelabel#1{\useafter\gdef{prelab@#1}}

\def\ifundefinedlabel#1\then{%
     \expandafter\ifx\csname lab@#1\endcsname\relax}
\def\useiflab#1\then{\csname iflab@#1\endcsname}

\def\prelabelcheck{{%
     \def\^^\##1{\useiflab{##1}\then\else\err@undefinedlabel{##1}\fi}%
     \@prelabellist}}

\message{equation numbering,}

\newcount\chapternum
\newcount\sectionnum
\newcount\subsectionnum
\newcount\equationnum
\newcount\subequationnum
\newcount\figurenum
\newcount\subfigurenum
\newcount\tablenum
\newcount\subtablenum
\newcount\defnum
\newcount\subdefnum
\newcount\thmnum
\newcount\subthmnum
\newcount\lemnum
\newcount\sublemnum

\newif\if@subeqncount
\newif\if@subfigcount
\newif\if@subtblcount
\newif\if@subdefcount
\newif\if@subthmcount
\newif\if@sublemcount

\def\newchapternum{\newsectionnum=\z@\@resetnum\chapternum}
\def\newsectionnum{\newsubsectionnum=\z@\@resetnum\sectionnum}
\def\newsubsectionnum{\newequationnum=\z@\newfigurenum=\z@\newtablenum=\z@
     \newdefnum=\z@\newthmnum=\z@\newlemnum=\z@
     \@resetnum\subsectionnum}
\def\newequationnum{\newsubequationnum=\z@\@resetnum\equationnum}
\def\newsubequationnum{\@resetnum\subequationnum}
\def\newfigurenum{\newsubfigurenum=\z@\@resetnum\figurenum}
\def\newsubfigurenum{\@resetnum\subfigurenum}
\def\newtablenum{\newsubtablenum=\z@\@resetnum\tablenum}
\def\newsubtablenum{\@resetnum\subtablenum}
\def\newdefnum{\newsubdefnum=\z@\@resetnum\defnum}
\def\newsubdefnum{\@resetnum\subdefnum}
\def\newthmnum{\newsubthmnum=\z@\@resetnum\thmnum}
\def\newsubthmnum{\@resetnum\subthmnum}
\def\newlemnum{\newsublemnum=\z@\@resetnum\lemnum}
\def\newsublemnum{\@resetnum\sublemnum}

\def\@resetnum#1{\global\advance#1by1 \edef\next{\the#1\relax}\global#1}

\newchapternum=0

\def\chapternumstyle#1{\@setnumstyle\chapternum{#1}}
\def\sectionnumstyle#1{\@setnumstyle\sectionnum{#1}}
\def\subsectionnumstyle#1{\@setnumstyle\subsectionnum{#1}}
\def\equationnumstyle#1{\@setnumstyle\equationnum{#1}}
\def\subequationnumstyle#1{\@setnumstyle\subequationnum{#1}%
     \ifblank\subequationnumstyle\then\global\@subeqncountfalse\fi
     \ignorespaces}
\def\figurenumstyle#1{\@setnumstyle\figurenum{#1}}
\def\subfigurenumstyle#1{\@setnumstyle\subfigurenum{#1}%
     \ifblank\subfigurenumstyle\then\global\@subfigcountfalse\fi
     \ignorespaces}
\def\tablenumstyle#1{\@setnumstyle\tablenum{#1}}
\def\subtablenumstyle#1{\@setnumstyle\subtablenum{#1}%
     \ifblank\subtablenumstyle\then\global\@subtblcountfalse\fi
     \ignorespaces}
\def\defnumstyle#1{\@setnumstyle\defnum{#1}}
\def\subdefnumstyle#1{\@setnumstyle\subdefnum{#1}%
     \ifblank\subdefnumstyle\then\global\@subdefcountfalse\fi
     \ignorespaces}
\def\thmnumstyle#1{\@setnumstyle\thmnum{#1}}
\def\subthmnumstyle#1{\@setnumstyle\subthmnum{#1}%
     \ifblank\subthmnumstyle\then\global\@subthmcountfalse\fi
     \ignorespaces}
\def\lemnumstyle#1{\@setnumstyle\lemnum{#1}}
\def\sublemnumstyle#1{\@setnumstyle\sublemnum{#1}%
     \ifblank\sublemnumstyle\then\global\@sublemcountfalse\fi
     \ignorespaces}

\def\heqnlabel{\newequationnum=\next
          \ifblank\subequationnumstyle\then
          \else\global\@subeqncounttrue
               \newsubequationnum=\@ne
          \fi}

\def\eqnlabel#1{%
     \if@subeqncount
          \newsubequationnum=\next
     \else\heqnlabel
     \fi
     \Label{#1}{\puteqnformat}(\puteqn{#1})%
     \ifdraft\rlap{\hskip.1in{\tt#1}}\fi}

\let\puteqn=\putlab

\def\putequation#1{\useafter\ifundefined{eqn@#1}\then
     \err@undefinedeqn{#1}\else\usename{eqn@#1}\fi}

\def\eqnseriesstyle#1{\gdef\@eqnseriesstyle{#1}}
\def\begineqnseries{\subequationnumstyle{\@eqnseriesstyle}%
     \defaultoption[]\@begineqnseries}
\def\@begineqnseries[#1]{\edef\@@eqnname{#1}}
\def\endeqnseries{\subequationnumstyle{blank}%
     \expandafter\ifnoarg\@@eqnname\then
     \else\Label\@@eqnname{\puteqnformat}%
     \fi
     \aftergroup\ignorespaces}

\def\figlabel#1{%
     \if@subfigcount
          \newsubfigurenum=\next
     \else\newfigurenum=\next
          \ifblank\subfigurenumstyle\then
          \else\global\@subfigcounttrue
               \newsubfigurenum=\@ne
          \fi
     \fi
     \Label{#1}{\putfigformat}\putfig{#1}%
   }

\let\putfig=\putlab

\def\figseriesstyle#1{\gdef\@figseriesstyle{#1}}
\def\beginfigseries{\subfigurenumstyle{\@figseriesstyle}%
     \defaultoption[]\@beginfigseries}
\def\@beginfigseries[#1]{\edef\@@figname{#1}}
\def\endfigseries{\subfigurenumstyle{blank}%
     \expandafter\ifnoarg\@@figname\then
     \else\Label\@@figname{\putfigformat}%
     \fi
     \aftergroup\ignorespaces}

\def\tbllabel#1{%
     \if@subtblcount
          \newsubtablenum=\next
     \else\newtablenum=\next
          \ifblank\subtablenumstyle\then
          \else\global\@subtblcounttrue
               \newsubtablenum=\@ne
          \fi
     \fi
     \Label{#1}{\puttblformat}\puttbl{#1}%
}

\let\puttbl=\putlab

\def\tblseriesstyle#1{\gdef\@tblseriesstyle{#1}}
\def\begintblseries{\subtablenumstyle{\@tblseriesstyle}%
     \defaultoption[]\@begintblseries}
\def\@begintblseries[#1]{\edef\@@tblname{#1}}
\def\endtblseries{\subtablenumstyle{blank}%
     \expandafter\ifnoarg\@@tblname\then
     \else\Label\@@tblname{\puttblformat}%
     \fi
     \aftergroup\ignorespaces}


\def\deflab#1{%
     \if@subdefcount
          \newsubdefnum=\next
     \else\newdefnum=\next
          \ifblank\subdefnumstyle\then
          \else\global\@subdefcounttrue
               \newsubdefnum=\@ne
          \fi
     \fi
     \Label{#1}{\putdefformat}\refdef{#1}%
}

\let\refdef=\putlab

\def\defseriesstyle#1{\gdef\@defseriesstyle{#1}}
\def\begindefseries{\subtablenumstyle{\@defseriesstyle}%
     \defaultoption[]\@begindefseries}
\def\@begindefseries[#1]{\edef\@@defname{#1}}
\def\enddefseries{\subdefnumstyle{blank}%
     \expandafter\ifnoarg\@@defname\then
     \else\Label\@@defname{\putdefformat}%
     \fi
     \aftergroup\ignorespaces}

\def\thmlab#1{%
     \if@subthmcount
          \newsubthmnum=\next
     \else\newthmnum=\next
          \ifblank\subthmnumstyle\then
          \else\global\@subthmcounttrue
               \newsubthmnum=\@ne
          \fi
     \fi
     \Label{#1}{\putthmformat}\refthm{#1}%
}

\let\refthm=\putlab

\def\thmseriesstyle#1{\gdef\@thmseriesstyle{#1}}
\def\beginthmseries{\subthmnumstyle{\@thmseriesstyle}%
     \defaultoption[]\@beginthmseries}
\def\@beginthmseries[#1]{\edef\@@thmname{#1}}
\def\endthmseries{\subthmstyle{blank}%
     \expandafter\ifnoarg\@@thmname\then
     \else\Label\@@thmname{\putthmformat}%
     \fi
     \aftergroup\ignorespaces}

\def\lemlab#1{%
     \if@sublemcount
          \newsublemnum=\next
     \else\newlemnum=\next
          \ifblank\sublemnumstyle\then
          \else\global\@sublemcounttrue
               \newsublemnum=\@ne
          \fi
     \fi
     \Label{#1}{\putlemformat}\reflem{#1}%
}

\let\reflem=\putlab

\def\lemseriesstyle#1{\gdef\@lemseriesstyle{#1}}
\def\beginlemseries{\sublemnumstyle{\@lemseriesstyle}%
     \defaultoption[]\@beginlemseries}
\def\@beginlemseries[#1]{\edef\@@lemname{#1}}
\def\endlemseries{\sublemnumstyle{blank}%
     \expandafter\ifnoarg\@@lemname\then
     \else\Label\@@lemname{\putlemformat}%
     \fi
     \aftergroup\ignorespaces}

\message{reference numbering,}

\newcount\referencenum \referencenum=0
\newcount\@@prerefcount \@@prerefcount=0
\newcount\@@thisref
\newcount\@@lastref
\newcount\@@loopref
\newcount\@@refseq
\newdimen\refnumindent
\let\@undefreflist=\empty

\def\referencenumstyle#1{\@setnumstyle\referencenum{#1}}

\def\referencestyle#1{\usename{@ref#1}}

\def\@refsequential{%
     \gdef\@refpredef##1{\global\advance\referencenum by\@ne
          \let\^^\=0\Label{##1}{\^^\{\the\referencenum}}%
          \useafter\gdef{ref@\the\referencenum}{{##1}{\undefinedlabelformat}}}%
     \gdef\@reference##1##2{%
          \ifundefinedlabel##1\then
          \else\def\^^\####1{\global\@@thisref=####1\relax}\putlab{##1}%
               \useafter\gdef{ref@\the\@@thisref}{{##1}{##2}}%
          \fi}%
     \gdef\endputreferences{%
          \loop\ifnum\@@loopref<\referencenum
                    \advance\@@loopref by\@ne
                    \expandafter\expandafter\expandafter\@printreference
                         \csname ref@\the\@@loopref\endcsname
          \repeat
          \par}}

\def\@refpreordered{%
     \gdef\@refpredef##1{\global\advance\referencenum by\@ne
          \additemR##1\to\@undefreflist}%
     \gdef\@reference##1##2{%
          \ifundefinedlabel##1\then
          \else\global\advance\@@loopref by\@ne
               {\let\^^\=0\Label{##1}{\^^\{\the\@@loopref}}}%
               \@printreference{##1}{##2}%
          \fi}
     \gdef\endputreferences{%
          \def\^^\####1{\useiflab{####1}\then
               \else\reference{####1}{\undefinedlabelformat}\fi}%
          \@undefreflist
          \par}}

\def\beginprereferences{\par
     \def\reference##1##2{\global\advance\referencenum by1\@ne
          \let\^^\=0\Label{##1}{\^^\{\the\referencenum}}%
          \useafter\gdef{ref@\the\referencenum}{{##1}{##2}}}}
\def\endprereferences{\global\@@prerefcount=\the\referencenum\par}

\def\beginputreferences{\par
     \refnumindent=\z@\@@loopref=\z@
     \loop\ifnum\@@loopref<\referencenum
               \advance\@@loopref by\@ne
               \setbox\z@=\hbox{\referencenum=\@@loopref
                    \referencenumformat\enskip}%
               \ifdim\wd\z@>\refnumindent\refnumindent=\wd\z@\fi
     \repeat
     \putreferenceformat
     \@@loopref=\z@
     \loop\ifnum\@@loopref<\@@prerefcount
               \advance\@@loopref by\@ne
               \expandafter\expandafter\expandafter\@printreference
                    \csname ref@\the\@@loopref\endcsname
     \repeat
     \let\reference=\@reference}

\def\@printreference#1#2{\ifx#2\undefinedlabelformat\err@undefinedref{#1}\fi
     \noindent\ifdraft\rlap{\hskip\hsize\hskip.1in \tt#1}\fi
     \llap{\referencenum=\@@loopref\referencenumformat\enskip}#2\par}

\def\reference#1#2{{\par\refnumindent=\z@\putreferenceformat\noindent#2\par}}

\def\putref#1{\trim#1\to\@@refarg
     \expandafter\ifnoarg\@@refarg\then
          \toks@={\relax}%
     \else\@@lastref=-\@m\def\@@refsep{}\def\@more{\@nextref}%
          \toks@={\@nextref#1,,}%
     \fi\the\toks@}
\def\@nextref#1,{\trim#1\to\@@refarg
     \expandafter\ifnoarg\@@refarg\then
          \let\@more=\relax
     \else\ifundefinedlabel\@@refarg\then
               \expandafter\@refpredef\expandafter{\@@refarg}%
          \fi
          \def\^^\##1{\global\@@thisref=##1\relax}%
          \global\@@thisref=\m@ne
          \setbox\z@=\hbox{\putlab\@@refarg}%
     \fi
     \advance\@@lastref by\@ne
     \ifnum\@@lastref=\@@thisref\advance\@@refseq by\@ne\else\@@refseq=\@ne\fi
     \ifnum\@@lastref<\z@
     \else\ifnum\@@refseq<\thr@@
               \@@refsep\def\@@refsep{,}%
               \ifnum\@@lastref>\z@
                    \advance\@@lastref by\m@ne
                    {\referencenum=\@@lastref\putrefformat}%
               \else\undefinedlabelformat
               \fi
          \else\def\@@refsep{--}%
          \fi
     \fi
     \@@lastref=\@@thisref
     \@more}

\message{streaming,}

\newif\ifstreaming

\def\streamto{\defaultoption[\jobname]\@streamto}
\def\@streamto[#1]{\global\streamingtrue
     \immediate\write\sixt@@n{--> Streaming to #1.str}%
     \newwrite\streamout\immediate\openout\streamout=#1.str }

\def\streamfrom{\defaultoption[\jobname]\@streamfrom}
\def\@streamfrom[#1]{\newread\streamin\openin\streamin=#1.str
     \ifeof\streamin
          \expandafter\err@nostream\expandafter{#1.str}%
     \else\immediate\write\sixt@@n{--> Streaming from #1.str}%
          \let\@@labeldef=\gdef
          \ifstreaming
               \edef\@elc{\endlinechar=\the\endlinechar}%
               \endlinechar=\m@ne
               \loop\read\streamin to\@@scratcha
                    \ifeof\streamin
                         \streamingfalse
                    \else\toks@=\expandafter{\@@scratcha}%
                         \immediate\write\streamout{\the\toks@}%
                    \fi
                    \ifstreaming
               \repeat
               \@elc
               \input #1.str
               \streamingtrue
          \else\input #1.str
          \fi
          \let\@@labeldef=\xdef
     \fi}

\def\streamcheck{\ifstreaming
     \immediate\write\streamout{\pagenum=\the\pagenum}%
     \immediate\write\streamout{\footnotenum=\the\footnotenum}%
     \immediate\write\streamout{\referencenum=\the\referencenum}%
     \immediate\write\streamout{\chapternum=\the\chapternum}%
     \immediate\write\streamout{\sectionnum=\the\sectionnum}%
     \immediate\write\streamout{\subsectionnum=\the\subsectionnum}%
     \immediate\write\streamout{\equationnum=\the\equationnum}%
     \immediate\write\streamout{\subequationnum=\the\subequationnum}%
     \immediate\write\streamout{\figurenum=\the\figurenum}%
     \immediate\write\streamout{\subfigurenum=\the\subfigurenum}%
     \immediate\write\streamout{\tablenum=\the\tablenum}%
     \immediate\write\streamout{\subtablenum=\the\subtablenum}%
     \immediate\closeout\streamout
     \fi}


\def\err@badtypesize{%
     \errhelp={The limited availability of certain fonts requires^^J%
          that the base type size be 10pt, 12pt, or 14pt.^^J}%
     \errmessage{--> Illegal base type size}}

\def\err@badsizechange{\immediate\write\sixt@@n
     {--> Size change not allowed in math mode, ignored}}

\def\err@sizetoolarge#1{\immediate\write\sixt@@n
     {--> \noexpand#1 too big, substituting HUGE}}

\def\err@sizenotavailable#1{\immediate\write\sixt@@n
     {--> Size not available, \noexpand#1 ignored}}

\def\err@fontnotavailable#1{\immediate\write\sixt@@n
     {--> Font not available, \noexpand#1 ignored}}

\def\err@sltoit{\immediate\write\sixt@@n
     {--> Style \noexpand\sl not available, substituting \noexpand\it}%
     \it}

\def\err@bfstobf{\immediate\write\sixt@@n
     {--> Style \noexpand\bfs not available, substituting \noexpand\bf}%
     \bf}

\def\err@badgroup#1#2{%
     \errhelp={The block you have just tried to close was not the one^^J%
          most recently opened.^^J}%
     \errmessage{--> \noexpand\End{#1} doesn't match \noexpand\begin{#2}}}

\def\err@badcountervalue#1{\immediate\write\sixt@@n
     {--> Counter (#1) out of bounds}}

\def\err@extrafootnotemark{\immediate\write\sixt@@n
     {--> \noexpand\footnotemark command
          has no corresponding \noexpand\footnotetext}}

\def\err@extrafootnotetext{%
     \errhelp{You have given a \noexpand\footnotetext command without first
          specifying^^Ja \noexpand\footnotemark.^^J}%
     \errmessage{--> \noexpand\footnotetext command has no corresponding
          \noexpand\footnotemark}}

\def\err@labelredef#1{\immediate\write\sixt@@n
     {--> Label "#1" redefined}}

\def\err@badlabelmatch#1{\immediate\write\sixt@@n
     {--> Definition of label "#1" doesn't match value in \jobname.lab}}

\def\err@needlabel#1{\immediate\write\sixt@@n
     {--> Label "#1" cited before its definition}}

\def\err@undefinedlabel#1{\immediate\write\sixt@@n
     {--> Label "#1" cited but never defined}}

\def\err@undefinedeqn#1{\immediate\write\sixt@@n
     {--> Equation "#1" not defined}}

\def\err@undefinedref#1{\immediate\write\sixt@@n
     {--> Reference "#1" not defined}}

\def\err@nostream#1{%
     \errhelp={You have tried to input a stream file that doesn't exist.^^J}%
     \errmessage{--> Stream file #1 not found}}

\message{jyTeX initialization}

\everyjob{\immediate\write16{--> jyTeX version \fmtversion}%
     \edef\@@jobname{\jobname}%
     \edef\jobname{\@@jobname}%
     \settime
     \openin0=\jobname.lab
     \ifeof0
     \else\closein0
          \immediate\write16{--> Getting labels from file \jobname.lab}%
          \input\jobname.lab
     \fi}


%
     \^^\{\splittopskip}%
     \^^\{\maxdepth}%
     \^^\{\skip\topins}%
     \^^\{\skip\footins}%
     \^^\{\headskip}%
     \^^\{\footskip}}

\def\scalingskipslist{%
     \^^\{\p@renwd}%
     \^^\{\delimitershortfall}%
     \^^\{\nulldelimiterspace}%
     \^^\{\scriptspace}%
     \^^\{\jot}%
     \^^\{\normalbaselineskip}%
     \^^\{\normallineskip}%
     \^^\{\normallineskiplimit}%
     \^^\{\baselineskip}%
     \^^\{\lineskip}%
     \^^\{\lineskiplimit}%
     \^^\{\bigskipamount}%
     \^^\{\medskipamount}%
     \^^\{\smallskipamount}%
     \^^\{\parskip}%
     \^^\{\parindent}%
     \^^\{\abovedisplayskip}%
     \^^\{\belowdisplayskip}%
     \^^\{\abovedisplayshortskip}%
     \^^\{\belowdisplayshortskip}%
     \^^\{\abovechapterskip}%
     \^^\{\belowchapterskip}%
     \^^\{\abovesectionskip}%
     \^^\{\belowsectionskip}%
     \^^\{\abovesubsectionskip}%
     \^^\{\belowsubsectionskip}}


\def\twoupsetup{
     \topmargin=.75in
     \leftmargin=.5in
     \vsize=6.9in
     \hsize=4.75in
     \fullhsize=10in
     \let\draft=\relax}


\chapterstyle{left}                              
\chapternumstyle{blank}                          
\def\chapterbreak{\newpage}                      
\abovechapterskip=0pt                            
\belowchapterskip=1.5\baselineskip               
     plus.38\baselineskip minus.38\baselineskip
\def\chapternumformat{\numstyle\chapternum.}     

\sectionstyle{left}                              
\sectionnumstyle{blank}                          
\def\sectionbreak{\vskip0pt plus4\baselineskip\penalty-100
     \vskip0pt plus-4\baselineskip}              
\abovesectionskip=1.5\baselineskip               
     plus.38\baselineskip minus.38\baselineskip
\belowsectionskip=\the\baselineskip              
     plus.25\baselineskip minus.25\baselineskip
\def\sectionnumformat{
     \ifblank\chapternumstyle\then\else\numstyle\chapternum.\fi
     \numstyle\sectionnum.}

\subsectionstyle{left}                           
\subsectionnumstyle{blank}                       
\def\subsectionbreak{\vskip0pt plus4\baselineskip\penalty-100
     \vskip0pt plus-4\baselineskip}              
\abovesubsectionskip=\the\baselineskip           
     plus.25\baselineskip minus.25\baselineskip
\belowsubsectionskip=.75\baselineskip            
     plus.19\baselineskip minus.19\baselineskip
\def\subsectionnumformat{
     \ifblank\chapternumstyle\then\else\numstyle\chapternum.\fi
     \ifblank\sectionnumstyle\then\else\numstyle\sectionnum.\fi
     \numstyle\subsectionnum.}


\def\undefinedlabelformat{$\bullet$}             


\equationnumstyle{arabic}                        
\subequationnumstyle{blank}                      
\figurenumstyle{arabic}                          
\subfigurenumstyle{blank}                        
\tablenumstyle{arabic}                           
\subtablenumstyle{blank}                         
\defnumstyle{arabic}                             
\subdefnumstyle{blank}                           
\thmnumstyle{arabic}                             
\subthmnumstyle{blank}                           
\lemnumstyle{arabic}                             
\sublemnumstyle{blank}                           

\eqnseriesstyle{alphabetic}                      
\figseriesstyle{alphabetic}                      
\tblseriesstyle{alphabetic}                      
\defseriesstyle{alphabetic}                      
\thmseriesstyle{alphabetic}                      
\lemseriesstyle{alphabetic}                      

\def\puteqnformat{\hbox{
     \ifblank\chapternumstyle\then\else\numstyle\chapternum.\fi
     \ifblank\sectionnumstyle\then\else\numstyle\sectionnum.\fi
     \ifblank\subsectionnumstyle\then\else\numstyle\subsectionnum.\fi
     \numstyle\equationnum
     \numstyle\subequationnum}}
\def\putfigformat{\hbox{
     \ifblank\chapternumstyle\then\else\numstyle\chapternum.\fi
     \ifblank\sectionnumstyle\then\else\numstyle\sectionnum.\fi
     \ifblank\subsectionnumstyle\then\else\numstyle\subsectionnum.\fi
     \numstyle\figurenum
     \numstyle\subfigurenum}}
\def\puttblformat{\hbox{
     \ifblank\chapternumstyle\then\else\numstyle\chapternum.\fi
     \ifblank\sectionnumstyle\then\else\numstyle\sectionnum.\fi
     \ifblank\subsectionnumstyle\then\else\numstyle\subsectionnum.\fi
     \numstyle\tablenum
     \numstyle\subtablenum}}
\def\putdefformat{\hbox{
     \ifblank\chapternumstyle\then\else\numstyle\chapternum.\fi
     \ifblank\sectionnumstyle\then\else\numstyle\sectionnum.\fi
     \ifblank\subsectionnumstyle\then\else\numstyle\subsectionnum.\fi
     \numstyle\defnum
     \numstyle\subdefnum}}
\def\putthmformat{\hbox{
     \ifblank\chapternumstyle\then\else\numstyle\chapternum.\fi
     \ifblank\sectionnumstyle\then\else\numstyle\sectionnum.\fi
     \ifblank\subsectionnumstyle\then\else\numstyle\subsectionnum.\fi
     \numstyle\thmnum
     \numstyle\subthmnum}}
\def\putlemformat{\hbox{
     \ifblank\chapternumstyle\then\else\numstyle\chapternum.\fi
     \ifblank\sectionnumstyle\then\else\numstyle\sectionnum.\fi
     \ifblank\subsectionnumstyle\then\else\numstyle\subsectionnum.\fi
     \numstyle\lemnum
     \numstyle\sublemnum}}


\referencestyle{sequential}                      
\referencenumstyle{arabic}                       
\def\putrefformat{\numstyle\referencenum}        
\def\referencenumformat{\numstyle\referencenum.} 
\def\putreferenceformat{
     \everypar={\hangindent=1em \hangafter=1 }%
     \def\\{\hfil\break\null\hskip-1em \ignorespaces}%
     \leftskip=\refnumindent\parindent=0pt \interlinepenalty=1000 }


\def\fmtversion{2.6M (June 1992)}


\def\ref#1{(\puteqn{#1})}
\def\label#1{\eqno\eqnlabel{#1}}
\font\bigboldfont=cmbx10 scaled \magstep2
\font\boldfont=cmbx10 scaled \magstep1
\def\displayhead#1{{\bigboldfont \leftline{#1}}
\vskip-10pt
\line{\hrulefill}}
\def\section#1{\ifblank\sectionnumstyle\then
          \else\newsectionnum=\next \fi
\displayhead{\ifblank\sectionnumstyle\then\else\sectionnumformat\ \fi#1}
     }
\def\subsection#1{\advance\subsectionnum by 1
  {\boldfont \leftline{\ifblank\sectionnumstyle\then\else\sectionnumformat\fi\number\subsectionnum\
        #1}} \vskip-10pt \line{\hrulefill}}
\def\appendix#1{\ifblank\sectionnumstyle\then
          \else\newsectionnum=\next \fi
\displayhead{Appendix
    \ifblank\sectionnumstyle\then\else\sectionnumformat\ \fi#1}
     }


\catcode`\@=12

\def\sn{\smallskip\noindent}
\def\mn{\medskip\noindent}
\def\bn{\bigskip\noindent}
\def\xv{\vec{x}}
\def\yv{\vec{y}}
\def\Sv{\vec{S}}
\def\Sz{S^z}
\def\Sp{S^{+}}
\def\Sm{S^{-}}

\def\abs#1{\vert #1 \vert}
\def\onehalf{{\textstyle {1 \over 2}}}
\def\Order{{\cal O}}
\def\Energy{{\cal E}}

\def\age{\,\raise2pt\hbox{$\mathop{>}\limits_{\raise 2pt
\hbox{$\sim$}}$}\,}
\def\ale{\,\raise2pt\hbox{$\mathop{<}\limits_{\raise 2pt
\hbox{$\sim$}}$}\,}
\chapternumstyle{blank}                           
\sectionnumstyle{arabic}                          
\def\cite#1{$\lbrack#1\rbrack$}
\def\bibitem#1{\parindent=9mm\item{\hbox to 7 mm{\cite{#1}\hfill}}}
\def\tabindents{\leftskip=2 true cm \rightskip=2 true cm}
\def\figindents{\leftskip=6 true pc \rightskip=2 true pc}
\def\Manou{1}
\def\OYA{2}
\def\Totsuka{3}
\def\CHP{4}
\def\CHPii{5}
\def\Tot{6}
\def\CaGy{7}
\def\poly{8}
\def\Korshunov{9}
\def\DoUi{10}
\def\Miyashita{11}
\def\MiNi{12}
\def\ChuGo{13}
\def\BLLP{14}
\def\SuMa{15}
\def\LoNo{16}
\def\YaMue{17}
\def\KoTa{18}
\def\ZhNi{19}
\def\TrSa{20}
\def\WWB{21}
\def\IAG{22}
\def\CoPe{23}
\def\HTM{24}
\def\HKSprep{25}
\def\BoFi{26}
\def\BoPa{27}
\def\Night{28}
\def\BiLe{29}
\def\SWW{30}
\def\Pert{31}
\def\WOH{32}
\def\WOHii{33}
\def\LeRu{34}
\def\Gluzman{35}
\def\SSS{36}
\def\HoPa{37}
\def\DzNe{38}
\def\PoTa{39}
\def\Popov{40}
\def\PopB{41}
\def\BaBra{42}
\def\SGMK{43}
\def\OHA{44}
\def\CHPzz{45}
\def\GJS{46}
\def\MBLW{47}
\def\MoSaKu{48}
\def\KuMo{49}
\def\GrSh{50}
\def\WEBSLLP{51}
\def\href#1#2{{#2}}
%
%
\font\large=cmbx10 scaled \magstep3
\font\bigf=cmr10 scaled \magstep2
\pageno=0
\def\folio{
\ifnum\pageno<1 \footline{\hfil} \else\number\pageno \fi}
\line{February 10, 1999 \hfill cond-mat/9902163}
\rightline{ETH-TH/99-02}
\vskip 1.5cm
\centerline{\large A Comparative Study of the Magnetization}
\vskip 5mm
\centerline{\large Process of Two-Dimensional Antiferromagnets}
\vskip 1.5cm
\centerline{\bigf A.\ Honecker\raise8pt\hbox{$\star$}}
\vskip 0.3cm
\centerline{\it Institut f\"ur Theoretische Physik, ETH-H\"onggerberg,
                CH--8093 Z\"urich, Switzerland}
\vskip 0.3cm
\centerline{\it honecker@itp.ethz.ch}
\vskip 2.0cm
\centerline{\bf Abstract}
\vskip 0.2truecm
\noindent
Plateaux in the magnetization curves of the square, triangular and
hexagonal lattice spin-$1/2$ XXZ antiferromagnet are investigated.
One finds a zero magnetization plateau (corresponding to a spin-gap)
on the square and hexagonal lattice with Ising-like anisotropies,
and a plateau with one third of the saturation magnetization
on the triangular lattice which survives a small amount of
easy-plane anisotropy. Here we start with transfer matrix computations
for the Ising limit and continue with series in the
XXZ-anisotropy for plateau-boundaries using the groundstates of
the Ising limit. The main focus is then a numerical computation
of the magnetization curves with anisotropies in the vicinity of the
isotropic situation. Finally, we discuss the universality class
associated to the asymptotic behaviour of the magnetization curve
close to saturation, as observed numerically in two and higher
dimensions.
\vfill
\noindent
\leftline{\hbox to 5 true cm{\hrulefill}}
\noindent
${}^{\star}$
{A Feodor-Lynen fellow of the Alexander von Humboldt-foundation}.
\eject
\noindent
\section{Introduction}
\mn
The discovery of high-$T_c$ superconductivity has revived interest
in two-dimensional Heisenberg antiferromagnets (for a review see
\cite{\Manou}), since the CuO$_2$ planes give rise to a good
realization of the $S=1/2$ square lattice antiferromagnet. Due to the
large coupling constants of the high-$T_c$ materials, the main
focus is on properties in zero or small external magnetic fields.
Nevertheless, it has also been progressively realized that
antiferromagnets exhibit interesting phenomena in strong
external magnetic fields, namely plateaux in their magnetization
curves at certain fractions of the saturation value of the magnetization.
\mn
In one dimension, the appearance of plateaux in magnetization
curves is by now rather well understood in terms of a quantization
condition on the magnetization that involves the volume of a
translationally invariant unit cell \cite{\OYA-\poly}. Actually,
if the interaction inside finite clusters of spins is large with
respect to the other interactions, the appearance of plateaux is
governed by the volume of such a cluster irrespective of the
dimension \cite{\poly}. However, the simplest two-dimensional
systems have equal coupling constants, and then it is less clear
what determines the appearance of plateaux in magnetization curves.
One well-known example of such plateaux in two dimensions is
a plateau at one third of the saturation magnetization in
the triangular lattice antiferromagnet. There is in fact a number
of theoretical studies of the magnetization process of two-dimensional
triangular antiferromagnets (see \cite{\Korshunov-\SuMa}
for a selection) which are at least to some extent motivated
by the presence of this plateau or the more general feature
of frustration. The number of recent investigations of the magnetization
process of the square lattice antiferromagnet \cite{\LoNo-\TrSa}
still seems to be smaller, and for the hexagonal lattice we are aware
of just a single study of the Ising antiferromagnet in a magnetic
field \cite{\WWB}.
\mn
The magnetization plateau of the triangular lattice antiferromagnet
can be observed experimentally in a variety of materials (see e.g.\
\cite{\IAG} for recent rather clear examples and \cite{\CoPe}
for a review of experimental facts about triangular lattice
antiferromagnets). The present investigation was in fact originally
motivated by a magnetization experiment on the stacked triangular
lattice antiferromagnet CsCuCl$_3$ \cite{\HTM} which
shows a plateau in the magnetization curve at
one third of the saturation value if the field is applied perpendicular
to the $c$-axis. In many cases and in particular in CsCuCl$_3$,
the spin is carried by a Cu$^{2+}$ ion, giving rise to
a spin $S=1/2$. Furthermore, at least in CsCuCl$_3$, anisotropy of the
interaction is important. This lead us to considering a spin-$1/2$
XXZ model and to investigate the effect of the XXZ-anisotropy
$\Delta$. The relevance of the results of the present paper for
CsCuCl$_3$ will be discussed elsewhere \cite{\HKSprep}.
\mn
The focus of the present work are general features, namely
in which situations magnetization plateaux arise in two dimensions
and the universality class of the transition to saturation.
These questions will be investigated by computing the zero-temperature
magnetization process of the $S=1/2$ XXZ model on the aforementioned
three lattice types, {\it i.e.}\ on a square, triangular
and hexagonal lattice. This problem is described by the
following Hamiltonian:
$$H = J \sum_{\langle \xv,\yv \rangle} \left\{
     \Delta \Sz_{\xv} \Sz_{\yv} + \onehalf \left(
     \Sp_{\xv} \Sm_{\yv} + \Sm_{\xv} \Sp_{\yv} \right) \right\}
   - h \sum_{\xv} \Sz_{\xv} \, ,
\label{hamOp2D}$$
where the $\Sv_{\xv}$ are spin-$1/2$ operators acting at
place $\xv$ and $h$ is a dimensionless magnetic field. The notation
$\langle \xv,\yv \rangle$ denotes neighbouring pairs on a lattice
whose total number of sites we denote by $V$.
The magnetization $\langle M \rangle$ is given by the
expectation value of the operator $M = {2 \over V} \sum_{\xv} \Sz_{\xv}$
where the prefactor is chosen in order to normalize the saturation
value to $\langle M \rangle = \pm 1$.
The magnetization operator $M$ commutes with the Hamiltonian \ref{hamOp2D}.
This leads to a technically useful simplification since it
allows one to relate all properties in a magnetic field $h$
to those at $h=0$ with a suitably fixed magnetization $\langle M \rangle$.
\mn
The plan of this paper is as follows: First, we compute magnetization
curves for the Ising model which is obtained from \ref{hamOp2D}
by dropping the $\Sp \Sm$ hopping matrix elements. Exact zero-temperature
groundstates are readily written down for all plateaux observed
in this limit. We then use this as an input to compute
perturbation series in $\Delta^{-1}$ for the gap of single-spin
excitations above these groundstates. Even though the boundaries
of plateaux are in general determined by multi-spin excitations,
the series for the single-spin excitations yield a guide in which
region the plateaux persist for $\Delta < \infty$. For general
XXZ-anisotropy $\Delta$, plateaux existing in the Ising limit
$\Delta = \infty$ may not just disappear, but further plateaux
could arise. This and the intrinsic limitations of a perturbative
approach necessitates a direct computation of the
magnetization process of the full quantum Hamiltonian \ref{hamOp2D},
in particular in the region where $\Delta$ is of order unity. The
bulk of the paper is therefore devoted to a numerical investigation
of \ref{hamOp2D} on all three lattice types which is much in the
spirit of classical work on single Heisenberg chains \cite{\BoFi,\BoPa}.
Finally we discuss the universality class associated to the asymptotic
behaviour of the magnetization curve close to saturation, {\it i.e.}\
for $\langle M \rangle \to 1$.
\bn
\section{The Ising antiferromagnet}
\mn
The simplest case where one can observe magnetization plateaux
in two dimensions is the Ising antiferromagnet. The latter
can be obtained from \ref{hamOp2D} by taking
the limit $\Delta \to \infty$ and rescaling $J \to J / \Delta$.
This yields the energy function
$$E(\{s_{\xv}\}) = {J \over 4} \sum_{\langle \xv,\yv \rangle}
     s_{\xv} s_{\yv}
   - {h \over 2} \sum_{\xv} s_{\xv} \, ,
\label{IsingEnerg}$$
where the $s_{\xv}$ are now Ising variables with $s_{\xv} = \pm 1$.
Note that here we use unusual conventions for $J$ and $h$ in order to
later simplify making contact with the XXZ Hamiltonian.
\mn
While in the bulk of the paper we restrict to zero temperature,
here it is actually useful to work at finite inverse temperature
$\beta$ in order to permit application of the transfer-matrix
method (see e.g.\ \cite{\Night}). The magnetization is then given by
$$\langle M \rangle = {
\sum_{\{s_{\xv}\}} s_{\yv} \; {\rm e}^{-\beta E(\{s_{\xv}\})}
\over
\sum_{\{s_{\xv}\}} {\rm e}^{-\beta E(\{s_{\xv}\})}
} \, .
\label{magIsing}$$
Note that here $M$ is no longer a conserved quantity and one can therefore
not drop the expectation values. Working with expectation values would
also be mandatory for an anisotropy axis which does not coincide with
the field direction, but this is not considered in the present paper.
\mn
In principle, one could directly write down zero-temperature groundstates
of \ref{IsingEnerg} and compute the values of $h$ where one of them
becomes preferred to others. Actually, a complete set of groundstates
is known since the two-dimensional Ising model in the presence of an
external magnetic field has already been studied some time ago on the
square lattice (see e.g.\ \cite{\BiLe}), the triangular lattice
(see e.g.\ \cite{\SWW}) and somewhat more recently on the hexagonal
lattice \cite{\WWB}.
\mn
However, since we later wish to build on the results for this limit,
we believe that it is still useful to summarize the results in a
uniform framework. We therefore evaluate \ref{magIsing} on a strip
using the transfer-matrix method \cite{\Night}, where we employ
periodic boundary conditions along the the short direction of the
strip and open ones in the long one. The site $\yv$ in \ref{magIsing}
is put at the center of the strip.
\mn
\centerline{
\psfig{figure=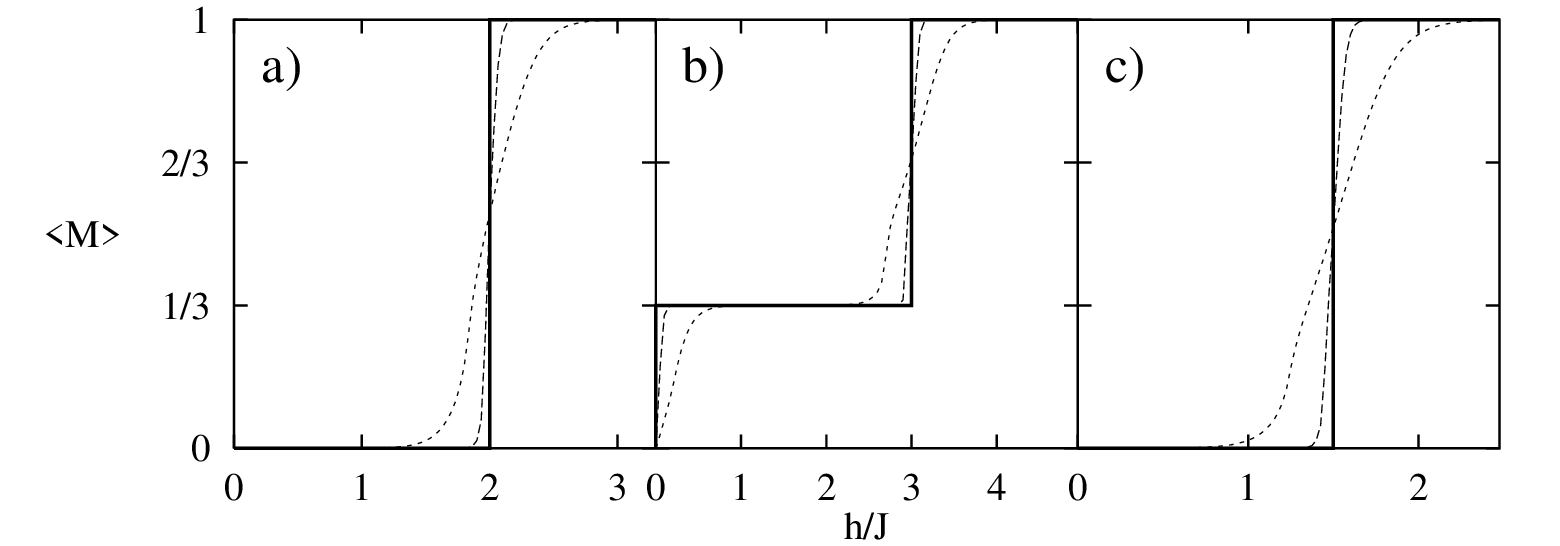,width=17 true cm}
}
\sn
{\par\noindent\figindents
{\bf Fig.\ 1:}
Magnetization curves of an Ising antiferromagnet on a $18 \times 1000$
a) square, b) triangular and c) hexagonal lattice.
The lines are for $J \beta = \infty$ (full), $J\beta = 40$ (long
dashes) and $J \beta = 8$ (short dashes).
\par\noindent}
\mn
Fig.\ 1 shows magnetization curves obtained in this manner on
a square, triangular and hexagonal lattice. The strip for Fig.\ 1 was always
chosen to be 1000 sites long and 18 sites wide. A computation
with the same length but half the width ({\it i.e.}\ 10 sites for the
square and hexagonal lattice, but 9 for the triangular one) leads
to values for $\langle M \rangle$  which differ at most by $10^{-2}$
from the ones shown. We may therefore expect that the magnetization
curves in Fig.\ 1 are basically indistinguishable from those of the
thermodynamic limit. The smallness of finite-size effects will be a
justification for using small system sizes later in the analysis of
the XXZ antiferromagnet, although then finite-size corrections should be
considered.
\mn
The interpretation of Fig.\ 1 is as follows. Both the square
and hexagonal lattices are bipartite lattices. Thus, their
zero-temperature groundstate at sufficiently small fields
is given by all spins pointing up on one sublattice
and pointing down on the other. If the magnetic field becomes
large enough to make it favourable to align one further
spin along the field, so it becomes for all and one has
a sharp transition from the unmagnetized antiferromagnetic
groundstate to a fully magnetized ferromagnetic one. The
transition field is readily computed to equal $h = 2 J$
or $h = {3 \over 2} J$ for the square and hexagonal lattice,
respectively (actually, it is easy to see that the transition
to a fully magnetized state takes place at $h={z \over 2} J$,
where $z$ is the coordination number of the lattice).
\mn
The situation is slightly different for the
triangular lattice where each plaquette is frustrated.
This lattice has three sublattices and energy is minimized
at small fields by aligning all spins in two of them along the
field and the ones in the third in the opposite direction. Increasing
the field it becomes again favourable for all the latter spins
simultaneously to align also along the field. Thus one obtains
a sharp first-order transition from a state with $\langle M \rangle = 1/3$
to the fully magnetized one at $h = 3 J$.
\mn
These transitions are sharp only at zero temperature ($\beta = \infty$)
and smoothed out by thermal fluctuations otherwise. Since such a smoothing
effect of finite temperature is generic, we will restrict
from now on to zero temperature after having illustrated
this effect in the case of the Ising model.
\bn
\section{Expansions around the Ising limit}
\mn
After this illustrative study of the Ising antiferromagent we now turn
to the XXZ model \ref{hamOp2D}. We use the groundstates
described in the previous section to expand the gap of single-spin
excitations in powers of $\Delta^{-1}$.
\mn
Since we wish to cover a variety of cases, it is convenient
to use a simple but general method for higher order
series expansions of a quantum mechanical system which
is summarized e.g.\ in Section 3 of \cite{\Pert}. This should be
sufficient to obtain an overview, but could certainly be extended
to higher orders using more sophisticated cluster expansions if
this should turn out to be desireable for concrete applications. 
\mn
For the {\it square} and {\it hexagonal} lattice there is just
a plateau at $\langle M \rangle = 0$. The lowest excitations
are those where a single spin is flipped with respect to the
antiferromagnetic groundstate. Due to the antiferromagnetic nature
of the groundstate, the first-order corrections to the energies
vanish and one finds a non-trivial dispersion only in second order
in $\Delta^{-1}$. Since both cases have been studied in detail in
\cite{\WOH} (see also references therein) and \cite{\WOHii},
respectively, we skip the details of the computation.
\mn
Analytical fourth-order expressions for this $S^z = 1$ gap are
given by
$$m = J \left(2 \Delta - {5 \over 3 \Delta}
+ {137 \over 432 \Delta^3} \right)
 + \Order(\Delta^{-4})
\label{gapSqLat}$$
for the {\it square} lattice, and
$$m = J \left({3 \over 2} \Delta - {15 \over 8 \Delta}
+ {295 \over 128 \Delta^3} \right)
 + \Order(\Delta^{-4})
\label{gapHexLat}$$
for the {\it hexagonal} lattice. Numerical versions of the
coefficients up to tenth order can be found in Table I of
\cite{\WOH} and \cite{\WOHii}, respectively. Applying a
na{\"\i}ve ratio test to the higher orders given in \cite{\WOH},
one concludes that the series \ref{gapSqLat} can be expected to
converge for $\Delta^{-2} \ale 1$.
\mn
Now we proceed with the {\it triangular lattice}. Recall that this has
three sublattices and the lowest-energy state with $\langle M \rangle = 1/3$
is obtained by aligning all spins on two sublattices up and the
ones on the third one down. The energy of this state at $h=0$
is found to be
$$E_{1/3} = - J V \left({\Delta \over 4} + {1 \over 4 \Delta}
          -{1 \over 8 \Delta^2} +{19 \over 320 \Delta^3}\right)
          + \Order(\Delta^{-4}) \, .
\label{gsTriagLat}$$
One possible excitation is obtained by flipping one further spin
up.
Using a Fourier transformation to lift the translational degeneracy,
one finds the dispersion relation
$$\eqalign{
\Energy_{+}(k_x,k_y) = &J \left[3 \Delta + {
\textstyle \cos{k_x} + \cos{k_y} - {3 \over 4} - 2
\left(\cos{k_x} + \cos{k_y}\right) \cos(k_x + k_y)
- \cos(k_x - k_y)
\over \Delta}
\right] \cr
&+ \Order(\Delta^{-2}) \, . \cr
}\label{dispTriagLatP}$$
{}From its minimum one obtains the gap for $\delta S^z = 1$
excitations above the $\langle M \rangle = 1/3$ plateau
$$\Energy_{+}(0,0) = \Energy_{+}\left( {2 \pi \over 3},{2 \pi \over 3}\right)
  = 3 J \Delta - {15 J \over 4 \Delta}
+ {75 J \over 16 \Delta^2} + {783 J \over 320 \Delta^3} + \Order(\Delta^{-4})
 \, .
\label{hc2TriagLat}$$
Now consider excitations with $S^z$ by $1$ smaller than on the
$\langle M \rangle = 1/3$ plateau state. There are two possibilities
to flip a spin down with respect to the plateau state
which are not related by translational symmetry.
In order to minimize the energy one has to take the
{\it difference} as a linear combination of these two possibilities.
Then one finds in a way similar to \ref{dispTriagLatP}
$$\eqalign{
\Energy_{-}(k_x,k_y) = -J &\left[
{1 \over 2} \left(\cos{k_x} + \cos{k_y} + \cos(k_x + k_y) \right) \right. \cr
& + {1 \over \Delta} \left(
{7 \over 8} + \cos(k_x + k_y) (\cos{k_x} + \cos{k_y} - \cos{k_x} \cos{k_y})
\right. \cr
&\left.\left.\quad
- {3 \over 4} (\cos(k_x + k_y) + \cos{k_x} + \cos{k_y})
+ \cos{k_x} \cos{k_y} \right)
\right] + \Order(\Delta^{-2}) \, . \cr
}\label{dispTriagLatM}$$
Its minimum determines the gap for $\delta S^z = -1$
excitations above the $\langle M \rangle = 1/3$ plateau and is given by
$$\Energy_{-}(0,0)
= - {3 J \over 2} - {5 J \over 8 \Delta}  + {73 J \over 32 \Delta^2}
 - {42787 J \over 11520 \Delta^3}
 + \Order(\Delta^{-4}) \, .
\label{hc1TriagLat}$$
These series will be compared to results of a numerical diagonalization
in the following sections.
\mn
We conclude this section by mentioning that the upper critical field
$h_{uc}$ at which the transition to a fully magnetized state takes
place is straightforwardly computed exactly if it is determined
by a single spin-flip. One finds
$$h_{uc} = \cases{
d J (\Delta + 1)           & ($d$-dimensional hypercubic lattice), \cr
3 J (\Delta + {1 \over 2}) & (triangular lattice), \cr
{3 \over 2} J (\Delta + 1) & (hexagonal lattice). \cr
}\label{valHuc}$$
For later use we have given here actually the value of
$h_{uc}$ for the $d$-dimensional hypercubic lattice -- the
result for the square lattice is given by its $d=2$ special case.
Note that the overall numerical factor is proportional to the
coordination number $z$ of the lattice. Furthermore, for the
bipartite lattices (hypercubic and hexagonal) one has
$h_{uc} = {z \over 2} J (\Delta + 1)$.
\bn
\section{Numerical diagonalization for the square lattice}
\mn
In this and the following sections we report results of
a numerical study of the magnetization process for XXZ
anisotropies around $\Delta = 1$. We have numerically calculated
the lowest eigenvalues as a function of the magnetization, wave
vectors and $\Delta$ on finite systems.
Below we just present the consequences for the
magnetization curves and skip the details of these standard
(but still CPU time intensive) computations.
\mn
We choose to present our results in terms of `magnetic phase diagrams'.
They show the projection of the more conventional magnetization
curves onto the axis of the magnetic field. The values of
$\langle M \rangle$ are thus assigned to different regions of the plot.
Since in this compact representation we save one axis, we
can display the variation of the magnetization
curves as a function of $\Delta$ in a single figure.
\mn
First we discuss the square lattice. This case
has also been studied with finite system diagonalizations
at $\Delta = 1$ on a $4 \times 4$ lattice \cite{\LoNo}
as well as more recently on larger lattices \cite{\YaMue}.
At $\Delta = 1$ also a second-order spin-wave investigation
has been performed \cite{\ZhNi}. For $\Delta > 1$ finite
system diagonalizations and quantum Monte Carlo simulations
have been carried out in \cite{\KoTa}. The system is therefore
well understood and provides a good check of our method.
Before presenting our results we recall from \cite{\KoTa} that
one finds a plateau with zero magnetization for $\Delta > 1$
whose boundary corresponds to a first-order phase transition,
{\it i.e.}\ in the thermodynamic limit the magnetization jumps
by a finite amount. This plateau disappears (its width tends to
zero) as $\Delta \to 1$.
\mn
Fig.\ 2 shows the magnetic phase diagram in the region with
$\Delta$ close to one on a square lattice of size $4 \times 6$,
{\it i.e.}\ with a volume of 24 spins. The thin full lines denote
boundaries of the magnetization plateaux $\langle M \rangle =
m/12$ ($m=0,\ldots,12$) which have to occur for this
system size. For other system sizes also other values for
$\langle M \rangle$ will be possible. Therefore, regions in
Fig.\ 2 where these lines are regularly spaced can be expected to
correspond to smooth transitions in the thermodynamic limit.
Bearing this in mind, one can clearly see a plateau with
magnetization $\langle M \rangle = 0$ for $\Delta \age 1.05$.
On the other hand, this plateau seems to be absent for
$\Delta \ale 0.95$. It should be noted that for $\Delta > 1.325$
the neighbour of $\langle M \rangle = 0$ has magnetization
$1/6$, {\it i.e.}\ here it is favourable to flip two
spins in the direction of the magnetic field rather than one.
This fact reflects the first-order nature of the transition.
In this region $\Delta > 1.325$, the energy required for an
excitation corresponding to a single flipped spin is shown by
a dashed line (otherwise the energy for such an excitation is
equal to the boundary of the $\langle M \rangle = 0$ plateau:
$h_{lc} = m$).
\mn
The extension to tenth order \cite{\WOH} of the series \ref{gapSqLat}
is shown by the the bold full line in Fig.\ 2. It should be
compared to the energy-gap for a single flipped spin which
is not always equal to the boundary of the $\langle M \rangle = 0$
plateau (in this case shown by the dashed line). In the
interval $1.1 < \Delta \le 1.5$, this series slightly overshoots
the finite-size data. In fact, the same can be observed in earlier
presentations \cite{\WOH,\KoTa} and therefore probably is
not due to a finite-size effect but missing higher-order
terms in the series which could be quite important since
we are looking at a region close to the limits of validity
of this series ($\Delta \ge 1$ should be its region of
convergence).
\mn
\centerline{
\psfig{figure=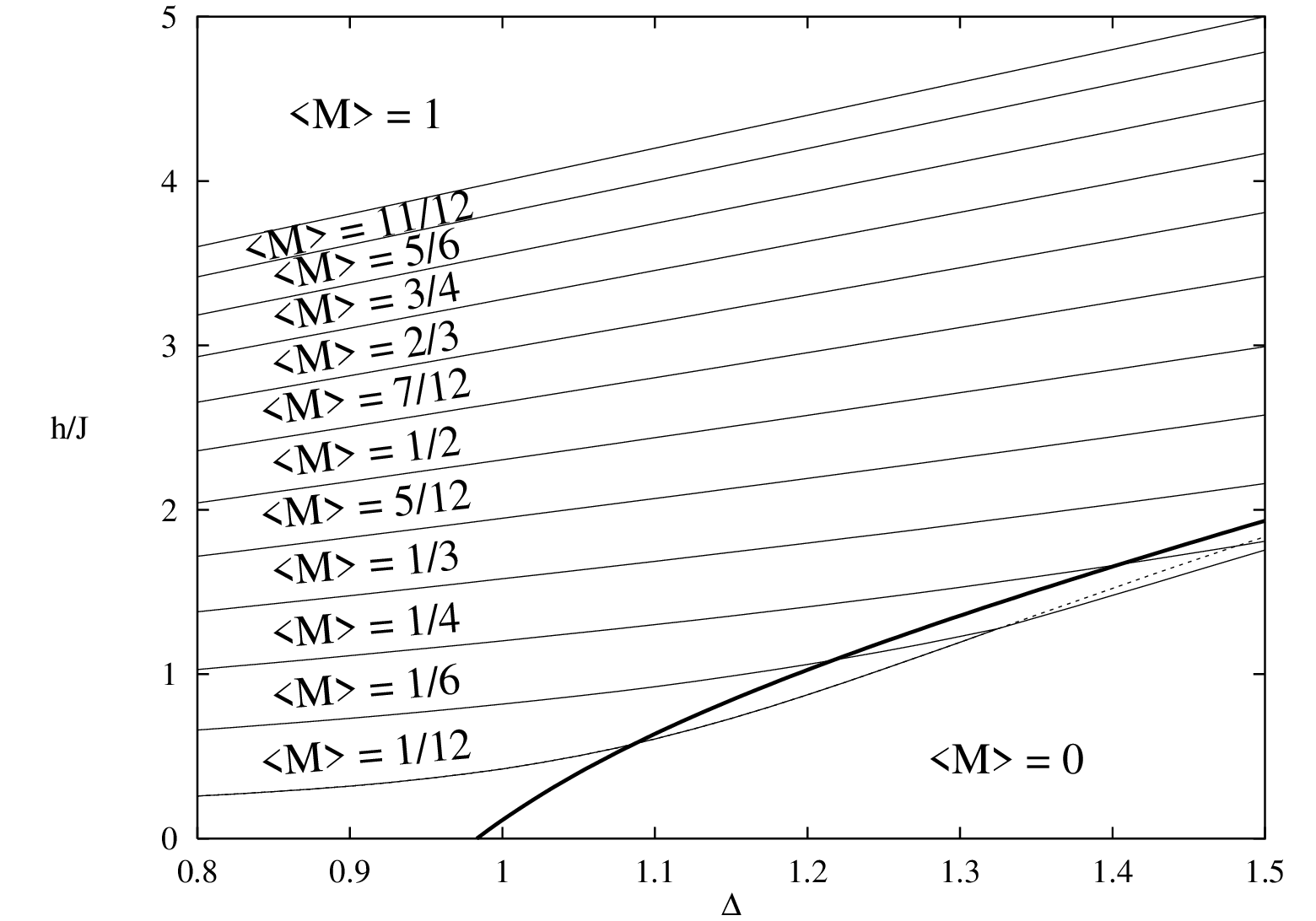,width=17 true cm}
}
\sn
{\par\noindent\figindents
{\bf Fig.\ 2:}
Magnetic phase diagram of the square lattice XXZ
antiferromagnet on a $4 \times 6$ lattice. The thin full
lines denote boundaries of areas with the values of the
magnetization indicated in the figure. The dashed line
is the spin-gap (see text).
The bold full line shows the extension to tenth order \cite{\WOH}
of the series \ref{gapSqLat} for the gap.
\par\noindent}
\mn
The fully magnetized state with $\langle M \rangle = 1$ gives
rise to a further trivial plateau in Fig.\ 2. The finite-size
data for its boundary agrees with the analytical result \ref{valHuc}
as it should.
\mn
Let us summarize our results for the square lattice: Using
a system of size $4 \times 6$ we are able to locate the end
of the zero magnetization plateau in the interval $0.95 \ale
\Delta \ale 1.05$ which is a reasonably good approximation to
the presumably exact value $\Delta = 1$. Furthermore, the
first-order nature of the transition reflects in the fact
that with this lattice size it is favourable to flip two spins
rather than one for $\Delta > 1.325$.
\vfill\eject
\section{The triangular lattice numerically}
\mn
The triangular lattice was already studied some time ago by
finite system diagonalizations \cite{\MiNi}. Now we can access
larger system sizes and look more carefully
at the dependence on $\Delta$ \footnote{${}^{1})$}{
The current record for finite system diagonalizations
on this type of lattice seems to be held with a volume
of 36 spins (see e.g.\ \cite{\LeRu,\BLLP}). Since we want to
vary the magnetization, wave vectors, $\Delta$ and still limit the
computational effort, we content ourselves with smaller
system sizes.
}. To obtain an overview, we first look at the $3 \times 6$ lattice,
even though precisely this lattice was already
studied in \cite{\MiNi} for $\Delta \in \{0.8, 1, 1.2, 2.5, 5\}$.
Fig.\ 3 shows our magnetic phase diagram for this case. The
magnetization curves at $\Delta = 1$, $1.2$ and $2.5$ of
\cite{\MiNi} correspond to sections through Fig.\ 3
and are in all cases in excellent agreement with our results.
\mn
\centerline{
\psfig{figure=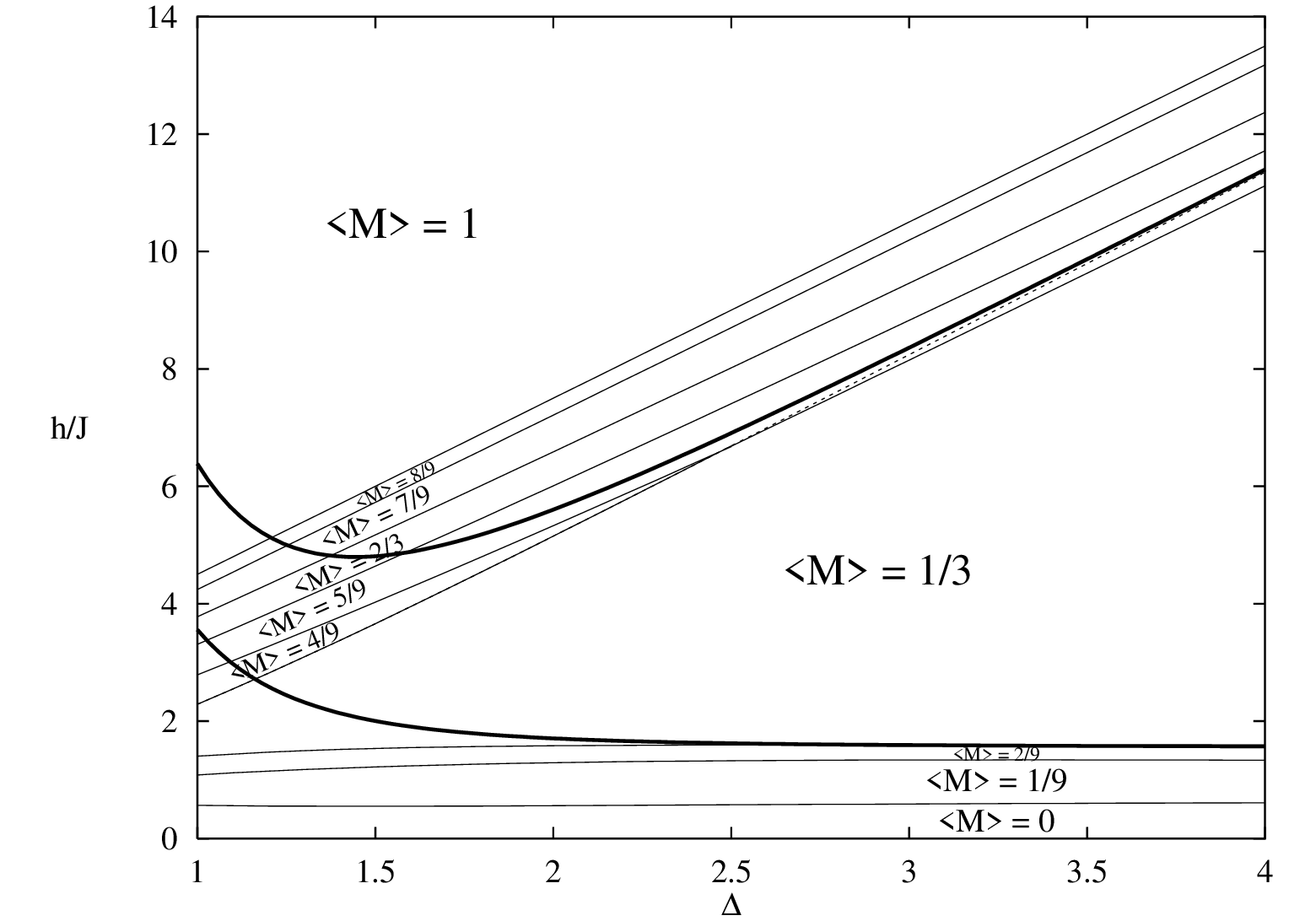,width=17 true cm}
}
\sn
{\par\noindent\figindents
{\bf Fig.\ 3:}
Magnetic phase diagram of the triangular lattice XXZ
antiferromagnet on a $3 \times 6$ lattice. The thin full
lines denote boundaries of areas with the values of the
magnetization indicated in the figure. The dashed line
is a single-spin excitation above the $\langle M \rangle = 1/3$
groundstate (see text).
The bold full lines show the series \ref{hc2TriagLat} and
\ref{hc1TriagLat}.
\par\noindent}
\mn
Here we observe precisely one non-trivial plateau with
$\langle M \rangle = 1/3$. Note that for $\Delta \ge 2.4$
it becomes favourable to flip two spins rather than one
at the upper boundary of this plateau. In this case, the
energy of an excitation corresponding to a single spin
flipped is shown by a dashed line while the finite-size data
for transitions between different groundstates is shown by thin full lines.
Based on our experience with the square lattice we take
this as an indication that the transition at the upper
boundary of this plateau becomes first order for
$\Delta \age 2$ while all other transitions are second order.
\mn
The series \ref{hc2TriagLat} and \ref{hc1TriagLat} for
the upper $h_{c_2} = \Energy_{+}(0,0)$ and lower boundaries
$h_{c_1} = -\Energy_{-}(0,0)$ of the $\langle M \rangle = 1/3$
plateau are shown by bold full lines. The former should
be compared to the dashed line, the latter to the appropriate
thin full line. The agreement is good for the right half of Fig.\ 3.
Since here we have less orders for the series than for the
square lattice (Fig.\ 2), it is not suprising that the agreement
in the region $\Delta$ close to one is less good.
Again, the finite-size data for the location of the
transition $\langle M \rangle \to 1$ agrees exactly
with \ref{valHuc}.
\mn
\centerline{
\psfig{figure=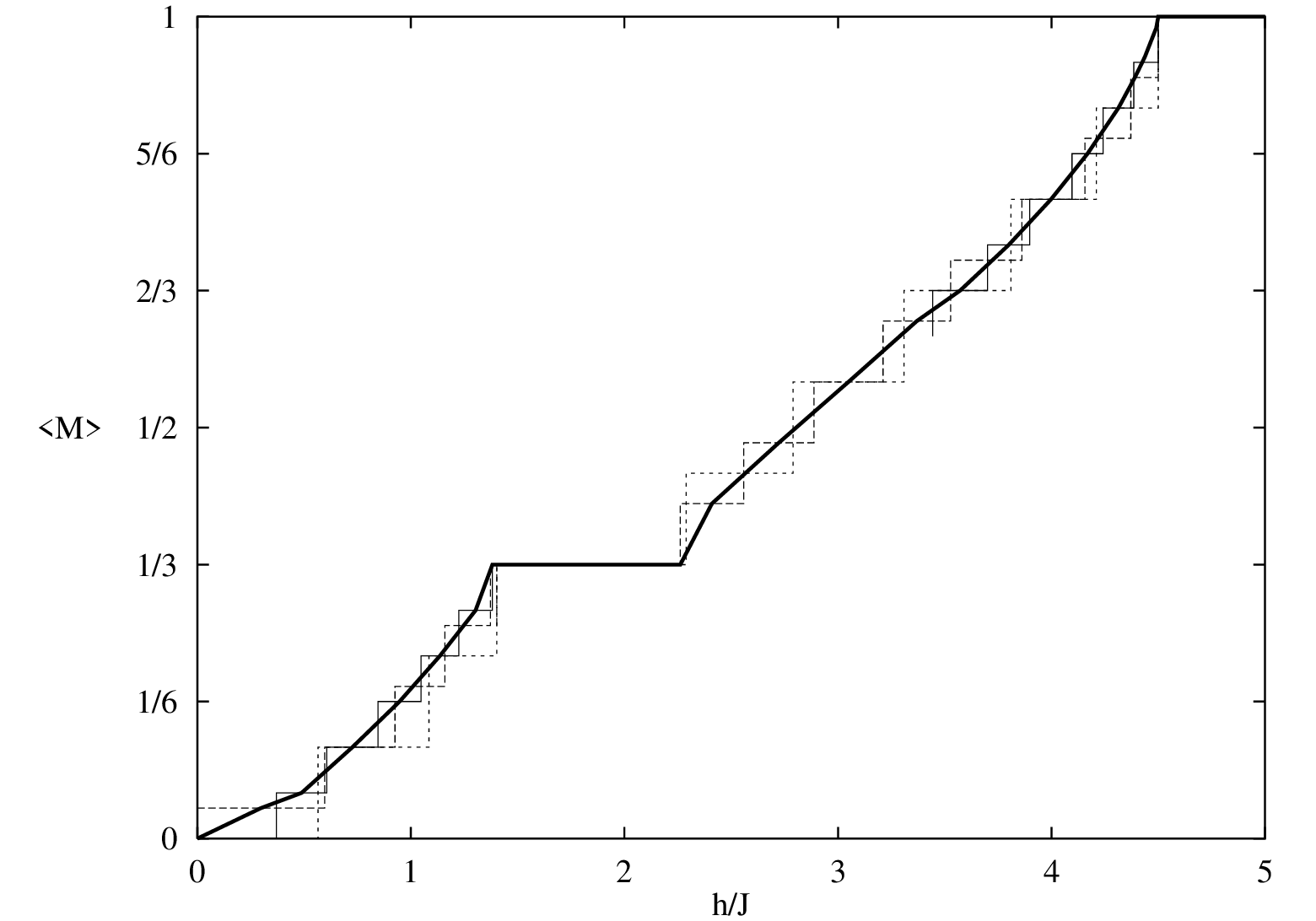,width=17 true cm}
}
\sn
{\par\noindent\figindents
{\bf Fig.\ 4:}
Magnetization curve of the spin-$1/2$ triangular lattice Heisenberg
antiferromagnet with $\Delta = 1$ on $3 \times 6$ (short dashes),
$V = 27$ (long dashes) and $V = 36$ (thin full line) lattices.
The curves with volume $V = 27$ and $V=36$ for $\langle M \rangle
\le 1/3$ are based on data of \cite{\BLLP}.
The bold full line is an extrapolation to the thermodynamic
limit (see text).
\par\noindent}
\mn
Now we examine the region $\Delta$ close to one in more detail.
First, we present the magnetization curve at $\Delta = 1$
in Fig.\ 4. The thin lines denote curves at three different system sizes.
Here, we have used results of \cite{\BLLP} to obtain the
parts with $\langle M \rangle \le 1/3$ of the magnetization
curves with $V=27$ and $V=36$. Our results overlap with
those of \cite{\BLLP} just at the lower boundary of the
$\langle M \rangle = 1/3$ plateau at a volume $V = 27$.
However, our geometry is different from that of \cite{\BLLP}:
In our computation the $V=27$ lattice has a smallest spatial extent
of just 3 sites while the configuration of \cite{\BLLP}
was designed to maximize the distance between boundaries.
This leads to a difference of $2\%$ between the results,
which is reasonably small in view of the small linear size.
\mn
In Fig.\ 4 one can clearly see a plateau with
$\langle M \rangle = 1/3$. The finite-size effects for
its boundaries are small \footnote{$^{2})$}{The first-order
spin-wave results for the boundaries of the $\langle M \rangle = 1/3$
plateau, $h_{c_1} \approx 1.248 J$ and $h_{c_2} \approx 2.145 J$
\cite{\ChuGo} are about $0.13 J$ smaller than the finite-size
diagonalization results.
}. One observes that finite-size effects
are also small for the midpoints of the steps in these curves,
as is well-known from one dimension \cite{\BoFi,\BoPa}.
Drawing a curve through these points for the largest available
system size (here we have used up to $V=225 = 15 \times 15$
in the vicinity of the upper critical field $h_{uc}$),
we therefore obtain the approximation to the thermodynamic limit
shown by the bold full line in Fig.\ 4.
\mn
\centerline{
\psfig{figure=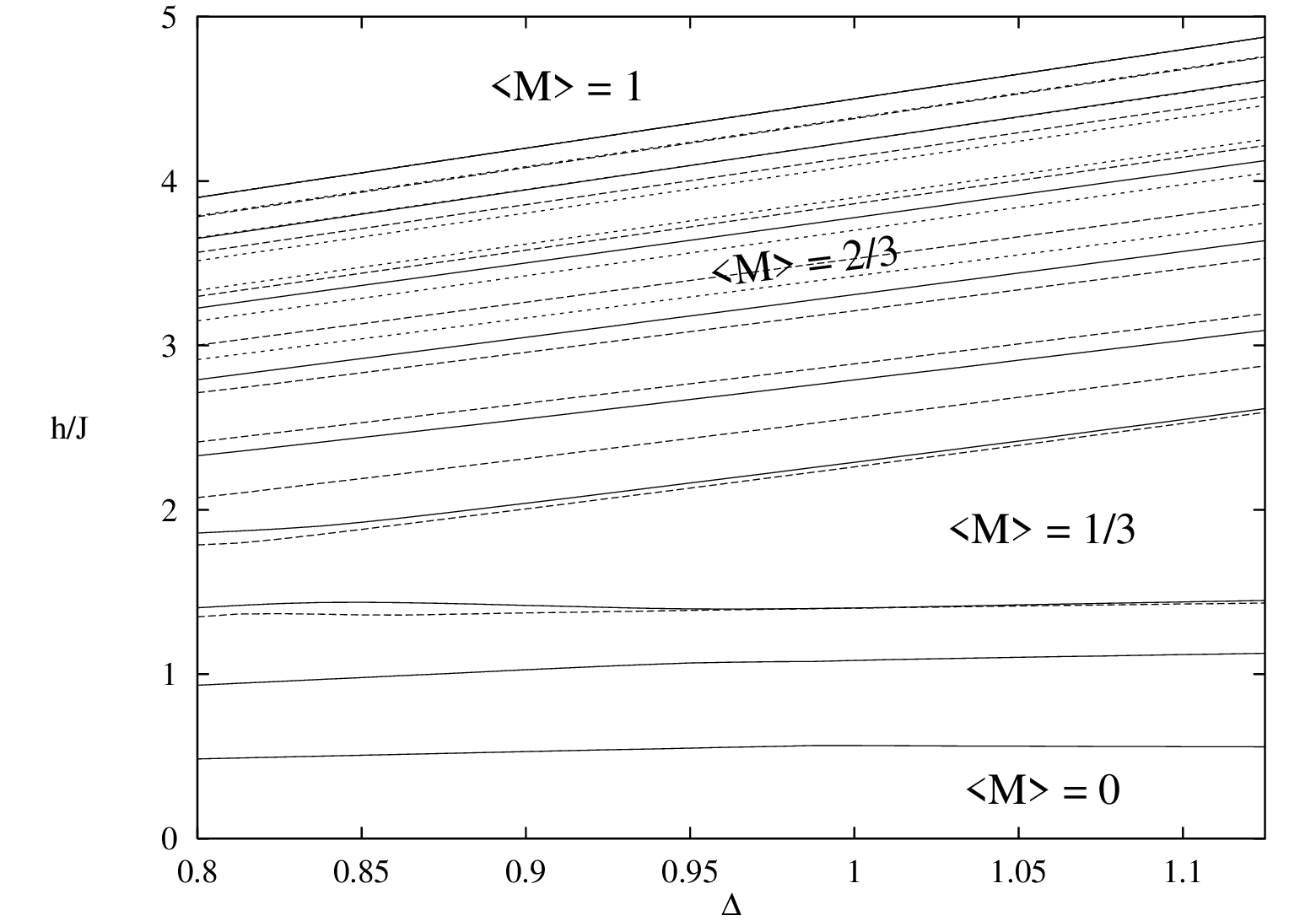,width=17 true cm}
}
\sn
{\par\noindent\figindents
{\bf Fig.\ 5:}
Magnetic phase diagram of the triangular lattice XXZ
antiferromagnet on a $3 \times 6$ (full line), $3 \times 9$
(long dashed line) and $6 \times 6$ (short dashed line) lattice.
\par\noindent}
\mn
Finally, in Fig.\ 5 we zoom in to the region $\Delta$ around one
of Fig.\ 3, using also bigger system sizes \footnote{${}^{3})$}{
We have not computed all transition lines for the larger lattices
sizes. We have omitted the ones for $\langle M \rangle < 1/3$ on the
$3 \times 9$ lattice and those with $\langle M \rangle < 2/3$ on the
$6 \times 6$ lattice.
} (Fig.\ 4 is a section
at $\Delta = 1$ through Fig.\ 5). The main motivation for taking
a closer look at this region comes from the observation in
\cite{\MiNi} that the $\langle M \rangle = 1/3$ plateau is
present at $\Delta = 1$, but does not seem to exist at $\Delta = 0.8$.
Inspecting Fig.\ 5 and paying attention to the finite-size
effects (in particular the difference between the $3 \times 6$
and the $3 \times 9$ lattice), one concludes that this plateau
presumably disappears somewhere in the region $\Delta \approx 0.85$.
\mn
We have also indicated the location of the $\langle M \rangle = 2/3$
plateau which is only realized for the $3 \times 6$ and the $6 \times 6$
lattice. However, the finite-size data provides no indication for it
to survive in the thermodynamic limit. So, it (and possible other
plateaux) are likely to be absent in the thermodynamic limit,
as is implied by the bold full line in Fig.\ 4.
\mn
In Fig.\ 5 we have omitted the series \ref{hc2TriagLat} and
\ref{hc1TriagLat} since already in Fig.\ 3 one can observe
notable deviations in this region of $\Delta$. Of course, the
result \ref{valHuc} is still valid.
\bn
\section{Numerical diagonalization for the hexagonal lattice}
\mn
While the square lattice is self-dual, the dual of the triangular
lattice is the hexagonal lattice. It is therefore interesting to
also investigate the magnetization process on the hexagonal lattice.
Fig.\ 6 shows the results of diagonalizations
on a $4 \times 6$ lattice. The region of small magnetization looks
qualitatively very similar to that of the square lattice in Fig.\ 2.
In particular, one can see an $\langle M \rangle = 0$ plateau
(corresponding to the gap) for $\Delta \age 1$. Inferring from
Fig.\ 6 that there is a first-order phase transition at the
boundary of this plateau may be speculative, but this
should also be expected on the grounds of universality, {\it i.e.}\
in the thermodynamic limit the transition at the boundary of the
$\langle M \rangle = 0$ plateau
for the hexagonal lattice should be in the same universality class
as that of the square lattice. The location of the ending-point
of the $\langle M \rangle = 0$ plateau is compatible with
$\Delta = 1$ (see also \cite{\WOHii})
which would be the same as that for the square lattice.
While in general one cannot use arguments based on universality
to locate a critical point, here the point $\Delta = 1$ is
distinguished by enhanced $su(2)$ symmetry and one may therefore
expect the closing of the gap exactly at $\Delta = 1$.
\mn
The tenth order version of \cite{\WOHii} of the series \ref{gapHexLat}
is shown by the bold full line in Fig.\ 6
and should be compared with the dashed line. The agreement is quite
good close to the right boundary of the figure and less good for
smaller values of $\Delta$. This is not surprising since the
series for the hexagonal lattice \cite{\WOHii} clearly converges
less well than that for the square lattice \cite{\WOH}.
\mn
As far as non-zero magnetizations are concerned, observe first
that the transition to the fully magnetized state does indeed take
place at the value of $h_{uc}$ given by \ref{valHuc}, thus giving
a crosscheck on our computations. More important are the candidates
for plateaux. In Fig.\ 6, the only plausible value for the
appearance of a non-trivial plateau is at $\langle M \rangle = 1/2$.
\mn
\centerline{
\psfig{figure=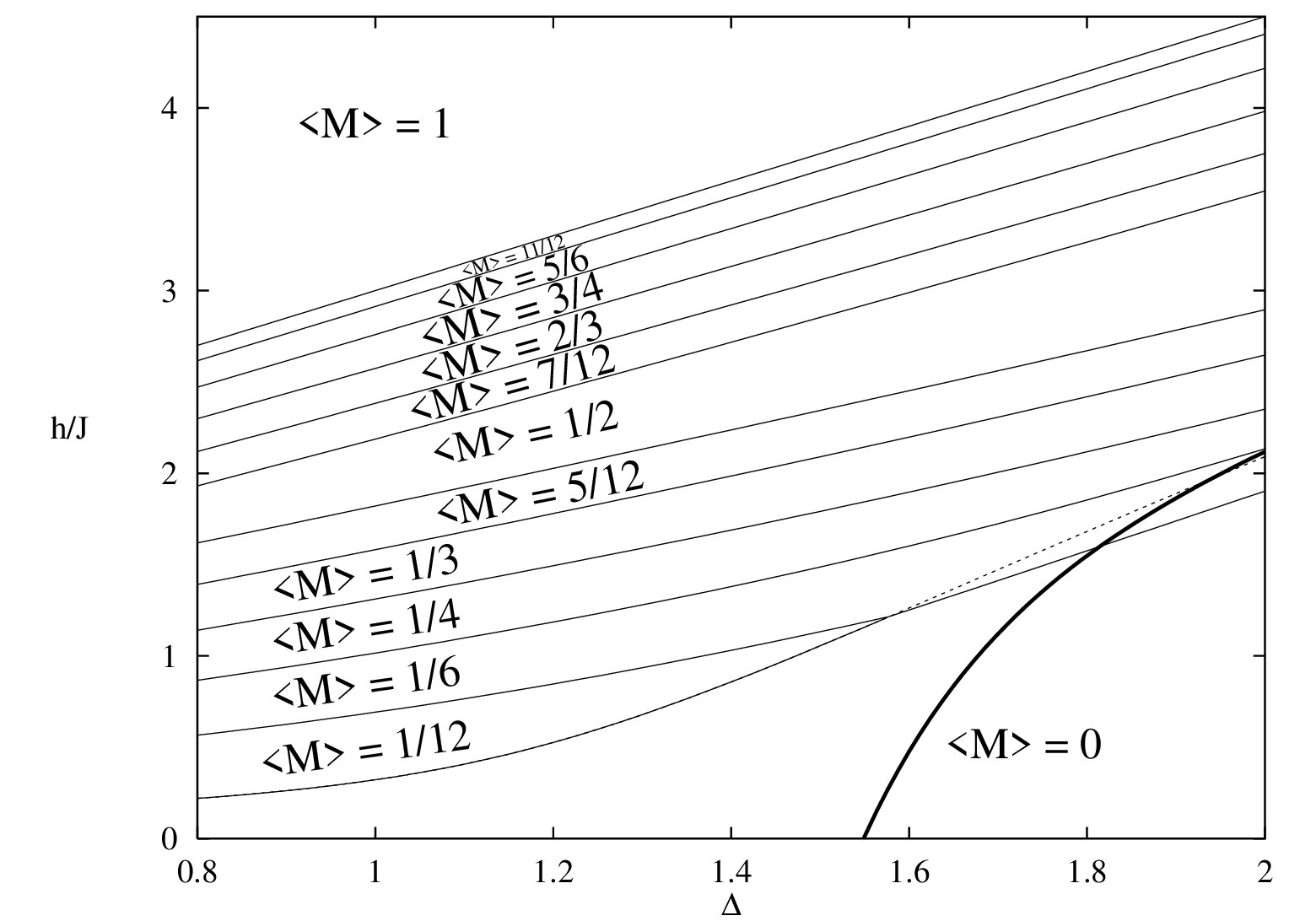,width=17 true cm}
}
\sn
{\par\noindent\figindents
{\bf Fig.\ 6:}
Same as Fig.\ 2, but for the hexagonal lattice.
The bold full line shows the extension to tenth order \cite{\WOHii}
of the series \ref{gapHexLat}.
\par\noindent}
\mn
It should be noted that the geometry corresponding to Fig.\ 6 can be
interpreted as a variant of the $N=4$-leg spin-ladder, and here plateaux are
expected at $\langle M \rangle = 0$ and $\langle M \rangle = 1/2$
\cite{\CHP,\CHPii}. So, in order to see whether the wide step
in Fig,\ 6 at $\langle M \rangle = 1/2$ is an intrinsic feature of
the hexagonal lattice, one should look at other system sizes, and in
particular larger strip widths. Since the large computational
effort makes a systematic investigation of the dependence on
$\Delta$ unfeasible, we have looked at the $6 \times 6$ lattice
for a few selected values of $\Delta$. The finite-size
magnetization curves for $\Delta = 1$ in Fig.\ 7 are representative
of the general case. In addition
to the data for the $4 \times 6$ lattice, we here also show
magnetization curves for the $4 \times 4$ lattice and $6 \times 6$
lattice (close to saturation, some larger lattice sizes have
also been considered, but are not explicitly shown).
\mn
With the data for the $6 \times 6$ lattice taken into account,
there is no indication anymore of an $\langle M \rangle = 1/2$ plateau.
The bold full line therefore shows an extrapolated magnetization curve
without plateaux, which as in Fig.\ 4 was obtained by connecting
the midpoints of the steps for the largest available system sizes.
This extrapolated curve shows that the $4 \times 6$ lattice is
still subject to substantial finite-size effects which at
$\langle M \rangle = 1/2$ conspire to suggest a plateau.
\mn
Also at $\Delta = 2$ there is no evidence either for a plateau
with $\langle M \rangle = 1/2$ on a $6 \times 6$ lattice
(note that here the corresponding candidate in Fig.\ 6 on a
$4 \times 6$ lattice is most pronounced). Thus, the final picture
for the hexagonal lattice is the same as for the
square lattice, {\it i.e.}\ the only plateau occurs for
$\Delta > 1$ at $\langle M \rangle = 0$. The main qualitative
difference between these two lattice types is in the finite-size
corrections which are much more important for the hexagonal lattice.
\mn
\centerline{
\psfig{figure=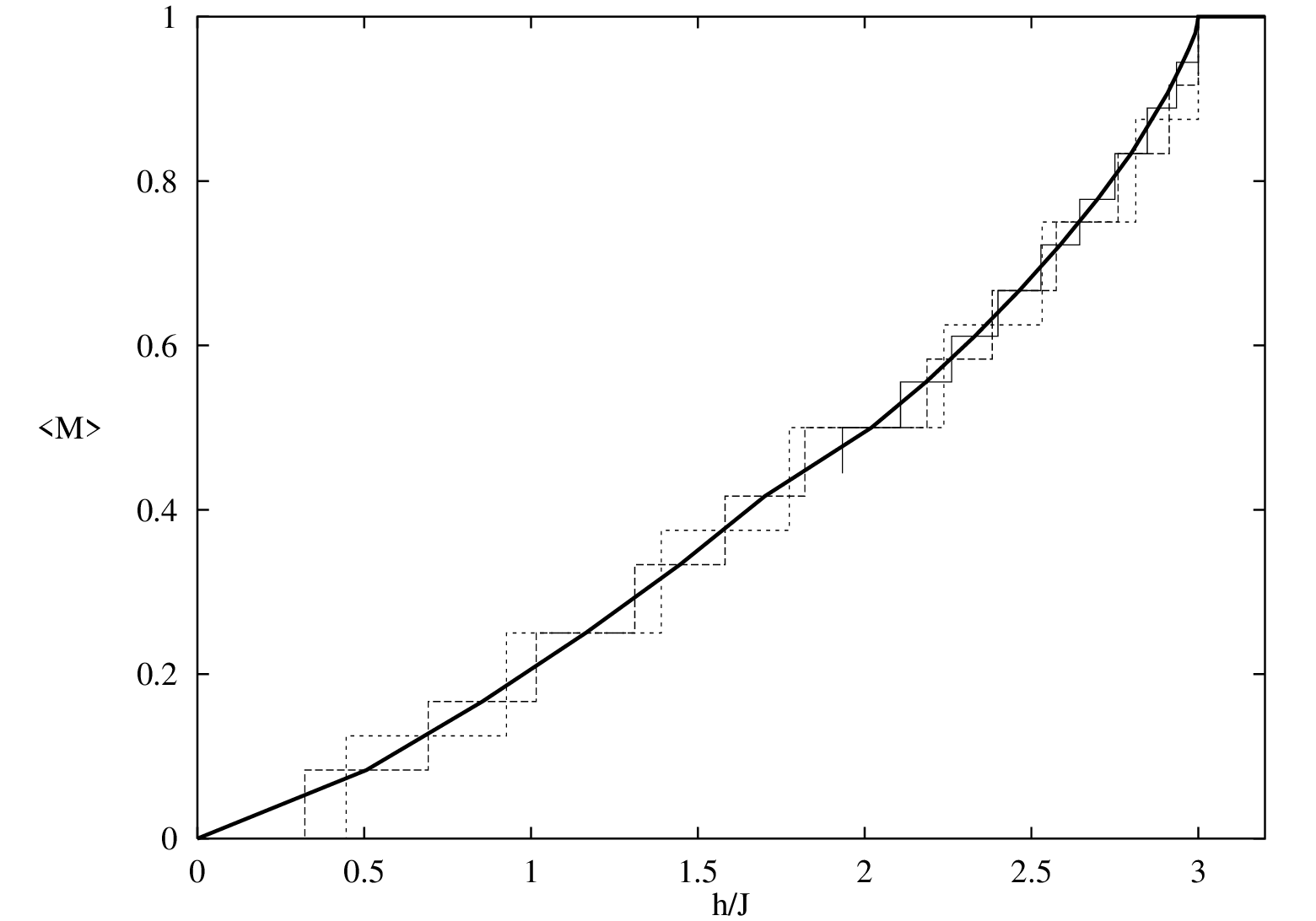,width=17 true cm}
}
\sn
{\par\noindent\figindents
{\bf Fig.\ 7:}
Magnetization curve of the hexagonal antiferromagnet with $\Delta = 1$.
The finite-size data is for $6 \times 6$ (thin full line),
$4 \times 6$ (long dashes) and $4 \times 4$ (short dashes) lattices,
respectively. As in Fig.\ 4, the bold full line is an extrapolation
to infinite volume.
\par\noindent}
\bn
\section{The transition to saturation}
\mn
Finally we look at the asymptotic behaviour the magnetization as
a function of the magnetic field close to saturation. One
possibility is that it reaches the upper critical field
$h_{uc}$ with a power law. Then one would introduce a critical
exponent $\mu$ via
$$1 - \langle M \rangle \sim
 \left({h_{uc} - h \over J}\right)^{\mu} \, .
\label{defMagExp}$$
A different possibility is a linear behaviour with a logarithmic
correction
$$1 - \langle M \rangle = a \left({h_{uc} - h \over J}\right)
     \ln\left( {b J \over h_{uc} - h} \right) \, .
\label{MagLog}$$
The latter has been argued in \cite{\ZhNi} to apply to the
square lattice antiferromagnet (see also \cite{\Gluzman,\SSS}).
\mn
To determine the functional form and estimate $\mu$ we use
the midpoint of the last step in the magnetization curve, as
was done for the one-dimensional case e.g.\ in \cite{\HoPa}.
Since the corresponding magnetization satisfies $1 - \langle M \rangle
= 2/V$, determination of the asymptotic behaviour of the
magnetization curve is equivalent to determining the asymptotic
finite-size behaviour of this last step. The particular choice
of the midpoint of the step is not relevant for the determination
of the exponent (or more generally, the functional form). Any other
choice such as the last corner would simply yield a
different prefactor. Values of such midpoints are given in Table 1
for all three lattice types at the isotropic point $\Delta = 1$.
\mn
Fitting this data for the square lattice to the form \ref{defMagExp},
we find a value of $\mu$ in the region $\mu \approx 0.83$.
However, the precise value increases if we use larger system
sizes for the fit, as is expected to be the case in the presence
of a logarithmic correction. Indeed, one obtains a better
fit if one uses \ref{MagLog} instead. Using all
data points for the square lattice in Table 1 we find
$a = 0.2505 \pm 0.0015$ and $b = 2.148 \pm 0.075$ where the
errors indicate the $1\sigma$ confidence interval of the fit.
So, the numerical data can be regarded as a confirmation of
the functional form \ref{MagLog} predicted by first-order spin-wave
theory \cite{\ZhNi}, though it is not surprising that the values
for the constants differ from the first-order spin-wave predictions
\cite{\ZhNi} which specialized to $S=1/2$ are $a = 1/(2 \pi)$, $b = \pi^2$.
In no case do we reproduce the simple linear behaviour
reported in \cite{\YaMue}. The crucial difference is probably
not that we employ exclusively system sizes which are larger than those
used in \cite{\YaMue}, but that the analysis in this reference
not only assumed the form \ref{defMagExp}, but also that $1/\mu$ is
an integer.
\mn
\centerline{
\hbox{
\vrule \hskip 1pt
\vbox{ \offinterlineskip
\def\tablespace{height2pt&\omit&&\omit&&\omit&&\omit&&\omit&&\omit&\cr}
\def\tablerule{ \tablespace
                \noalign{\hrule}
                \tablespace        }
\hrule
\halign{&\vrule#&
  \strut\hskip 4pt\hfil#\hfil\hskip 4pt\cr
height2pt&\multispan{3}&&\multispan{3}&&\multispan{3} & \cr
&\multispan{3} \hfil {\it square lattice} \hfil
  &&\multispan{3} \hfil {\it triangular lattice} \hfil
      &&\multispan{3} \hfil {\it hexagonal lattice} \hfil & \cr
height2pt&\multispan{3}&&\multispan{3}&&\multispan{3} & \cr
\noalign{\hrule}
\tablespace
& $V$ && $(h_{uc} - h)/J$
      && $V$ && $(h_{uc} - h)/J$
          && $V$ && $(h_{uc} - h)/J$ &\cr\tablerule
& $8 \times 8$ && $0.028767$
      && $6 \times 6$ && $0.057168$
          && $8 \times 8$ && $0.015650$ & \cr\tablespace
& $10 \times 10$ && $0.016407$
      && $9 \times 9$ && $0.022087$
          && $10 \times 10$ && $0.0089914$ & \cr\tablespace
& $12 \times 12$ && $0.010450$
      && $12 \times 12$ && $0.0094961$
          && $12 \times 12$ && $0.0057555$ & \cr\tablespace
& $14 \times 14$ && $0.0071704$
      && $15 \times 15$ && $0.0054802$
          && $14 \times 14$ && $0.0039638$ & \cr\tablespace
& $16 \times 16$ && $0.0051911$
      && $18 \times 18$ && $0.0035220$
          && $16 \times 16$ && $0.0028778$ & \cr\tablespace
& $18 \times 18$ && $0.0039129$
      && $21 \times 21$ && $0.0024338$
          && $18 \times 18$ && $0.0021741$ & \cr\tablespace
& $20 \times 20$ && $0.0030438$
      && $24 \times 24$ && $0.0017719$
          && $20 \times 20$ && $0.0016943$ & \cr\tablespace
& $22 \times 22$ && $0.0024282$
      && $27 \times 27$ && $0.0013419$
          && $22 \times 22$ && $0.0013540$ & \cr\tablespace
& $24 \times 24$ && $0.0019776$
      && $30 \times 30$ && $0.0010480$
          && $24 \times 24$ && $0.0011045$ & \cr\tablespace
& $26 \times 26$ && $0.0016386$
      && $33 \times 33$ && $0.00083883$
          && $26 \times 26$ && $0.00091571$ & \cr\tablespace
}
\hrule}\hskip 1pt \vrule}
}\par\noindent
{\tabindents
{\bf Table 1:} Midpoint of the last step before the upper
critical field $h_{uc}$ for two-dimensional antiferromagnets at $\Delta = 1$.
\par}
\mn
In the case of the triangular lattice, one should first discard
the data for he volumes $V = 6 \times 6$ and $V = 9 \times 9$
in order to obtain a smooth line. Then one again obtains the
same behaviour as for the square lattice: The exponent $\mu$
obtained by the fit \ref{defMagExp} has a very similar value and
moves in the same direction with increasing $V$ as for the
square lattice. Furthermore, one obtains a better fit with
\ref{MagLog} than with \ref{defMagExp}. Thus we conclude that
also the transition to saturation of the triangular lattice
antiferromagnet obeys \ref{MagLog}, where the constants are
now given by $a = 0.1495 \pm 0.0006$, $b=2.498 \pm 0.061$.
\mn
For the hexagonal lattice we can access
the same system sizes as for the square lattice. Thus, we can
directly compare the values
for the midpoints of the steps. One finds that those for the
hexagonal lattice differ from the ones for the square lattice by
a factor which approaches rapidly $\approx 0.56$ with
increasing system size. This means that the transitions to
saturation of the square and hexagonal lattice should belong to
the same universality class. Indeed, if we perform the same analysis
as for the square lattice, we find the same behaviour for the
hexagonal lattice. Just the constants for the fit \ref{MagLog}
are different: Now we have $a = 0.4338 \pm 0.0024$,
$b = 1.527 \pm 0.052$.
\mn
In summary, we find support for the asymptotic behaviour \ref{MagLog}
in all cases we have considered. This universal behaviour can be
understood in a way very similar to the DN-PT universal square root
in one dimension \cite{\DzNe,\PoTa}. One starts from the 
single-particle excitations. At the transition field one starts
to fill the lowest band of magnetic excitations. Generically,
the dispersion around such a minimum is quadratic, e.g.\
$E \sim \abs{\vec{k}}^2$. Then one needs to know how many
states are available below this value of $\abs{\vec{k}}$. In one
dimension, where the detailed nature of the excitations
does not matter, this number is proportional to $k$ as long as one
has an exclusion principle. Equivalently, in $d=1$ the number of states
available below a given energy $E$ is proportional to $\sqrt{E}$.
Since the magnetic field $h$ acts as a chemical potential and the
number of particles corresponds to the deviation of $\langle M \rangle$
from its critical value, this argument leads to the DN-PT universality class
in one dimension.
\mn
For hard-core bosons in two dimensions (which is the situation which
we consider here), interactions lead to a logarithmic correction
to the na\"{\i}ve dimensional analysis \cite{\Gluzman,\SSS,\Popov} (see also
Chapter 6 of \cite{\PopB}) and thus
to \ref{MagLog}. Note however that this argument is independent
of the details of the model under consideration. The crucial
ingredients are just that the fundamental excitations are
bosons with a quadratic dispersion around the minimum subject
to a repulsive interaction.
\mn
In three (and higher) dimensions we have Bose condensation. Therefore,
the magnetization curve of a hypercubic antiferromagnet in $d \ge 3$
should have a simple linear approach to saturation \cite{\BaBra}.
Since this is the behaviour found for classical spins, the
mean-field result for $\mu$ is in some sense exact in $d \ge 3$.
The effect of Bose condenstation which lowers $\mu$ substantially
below the value $d/2$ can be observed clearly in numerical
diagonalization of the isotropic (hyper)cubic antiferromagnet
(compare Table 2). In fact, if we fit the data in Table 2
to the form \ref{defMagExp}, we find $\mu \approx 0.95$ in
$d=3$ and $\mu \approx 0.98$ in $d=4$. Given the small linear
extent of the systems considered, this is in reasonable agreement
with the predicted value $\mu = 1$ without logarithmic
corrections.
\mn
Different values for $\mu$ can be obtained if the
form of the dispersion close to the minimum is not quadratic.
For example, for special values of parameters one could have
$E \sim \abs{\vec{k}}^4$ which in $d=1$ leads to $\mu = 1/4$.
The value $\mu = 1/4$ has been observed in the two-leg zig-zag ladder
\cite{\SGMK} and the biquadratic spin-1 chain \cite{\OHA} with
each a special value of the coupling constant. This is indeed explained
by the aforementioned change in the single-particle dispersion
(see \cite{\CHPzz} and \cite{\GJS}, respectively).
\mn
\centerline{
\hbox{
\vrule \hskip 1pt
\vbox{ \offinterlineskip
\def\tablespace{height2pt&\omit&&\omit&&\omit&&\omit&\cr}
\def\tablerule{ \tablespace
                \noalign{\hrule}
                \tablespace        }
\hrule
\halign{&\vrule#&
  \strut\hskip 4pt\hfil#\hfil\hskip 4pt\cr
height2pt&\multispan{3}&&\multispan{3} & \cr
&\multispan{3} \hfil $d=3$ \hfil
  &&\multispan{3} \hfil $d=4$ \hfil & \cr
height2pt&\multispan{3}&&\multispan{3} & \cr
\noalign{\hrule}
\tablespace
& $V$    && $(h_{uc} - h)/J$ && $V$    && $(h_{uc} - h)/J$ &\cr\tablerule
& $4^3$  && $0.064954$       && $4^4$  && $0.022792$ & \cr\tablespace
& $6^3$  && $0.017227$       && $6^4$  && $0.0043387$ & \cr\tablespace
& $8^3$  && $0.0068532$      && $8^4$  && $0.0013508$ & \cr\tablespace
& $10^3$ && $0.0033871$      && $10^4$ && $0.00054895$ & \cr\tablespace
& $12^3$ && $0.0019148$      && \omit  && \omit & \cr\tablespace
& $14^3$ && $0.0011859$      && \omit  && \omit & \cr\tablespace
& $16^3$ && $0.00078466$     && \omit  && \omit & \cr\tablespace
& $18^3$ && $0.00054581$     && \omit  && \omit & \cr\tablespace
}
\hrule}\hskip 1pt \vrule}
}\par\noindent
{\tabindents
{\bf Table 2:} Midpoint of the last step before the
transition to saturation for the cubic and $d=4$ hypercubic
antiferromagnet at $\Delta = 1$.
\par}
\bn
\section{Discussion and conclusions}
\mn
In the present paper we have observed the following plateaux in
magnetization curves: For the square and the hexagonal lattice,
{\it i.e.}\ the two bipartite lattices, we find an $\langle M \rangle = 0$
plateau for $\Delta > 1$. The transition at the boundary of this
plateau is likely to be always of first order. On the triangular
lattice one finds a plateau with $\langle M \rangle = 1/3$
for $\Delta \age 0.85$. The transitions at its boundary appear
to be second order for small enough anisotropies (at least for
$\Delta \ale 2$).
\mn
Recently plateaux with $\langle M \rangle = 1/2$ (and also
$\langle M \rangle = 0$) have been observed on the triangular lattice
with multi-spin interaction, in particular a four-spin interaction
in addition to the two-spin interaction discussed in the present
paper \cite{\MBLW,\MoSaKu}.
At least the $\langle M \rangle = 1/2$ plateau even survives
the classical limit \cite{\KuMo} \footnote{${}^{4})$}{In addition,
also a plateau with $\langle M \rangle = 1/3$ can be observed
in the classical model for a suitable choice of parameters
\cite{\KuMo}.}. This is to be contrasted with the $\langle M \rangle = 1/3$
plateau in the triangular lattice antiferromagnet which is absent
in the classical limit and arises only in first-order spin-wave
theory \cite{\ChuGo}.
\mn
While all the aforementioned plateaux with $\langle M \rangle \ne 0$
occur in frustrated systems, frustration is certainly not
a necessary ingredient for the appearance of non-trivial
plateaux. Consider for example a spin-$1/2$ $N$-layer square lattice
Heisenberg antiferromagnet (see \cite{\TrSa} for a detailed discussion
of the bilayer system in a magnetic field). In the limit where
the in-plane coupling tends to zero, this system decouples into
clusters of $N$ spins. Simply by counting the possible states
of these clusters, we conclude that plateaux exist with
$\langle M \rangle = -1$, $-1 + 2/N$, $\ldots$, $1 - 2/N$, $1$ at
least if the inter-plane coupling is much larger than the in-plane
coupling \cite{\CHP,\CHPii,\poly}.
However, frustration appears to favour the appearance of
further plateaux. Therefore, a bilayer triangular lattice may be an
interesting candidate for further study.
\mn
The triangular and hexagonal lattice are dual to each other and
thus share the same (local) point-symmetry. However, we have
seen that an antiferromagnet in a magnetic field behaves very
differently on them. The point-symmetry group therefore does not
appear to be of any relevance to the magnetization process.
On the other hand, the square and hexagonal lattice share the
property of being bipartite lattices and in fact gives rise
to very similar behaviour in the presence of a magnetic field.
\mn
In the classical or Ising limit it is clear that the appearance
of plateaux is related to the number of sublattices needed to
describe the full magnetization process. It is therefore likely
that some kind of translationally invariant unit cell will
control the appearance of plateaux also in higher dimensions.
The definition of such a unit cell in two dimensions is nevertheless
far less obvious than in one dimension, and some ambiguity is
also possible. To elucidate the situation further, it would
be useful to study a wider range of models and lattice types.
An important class of lattice types would be given by the eleven
Archimedean tilings (see e.g.\ Chapter 2 of \cite{\GrSh}) among
which we have considered the three monohedral ones. In particular,
it would be interesting to investigate the magnetization process
of the $S=1/2$ Heisenberg model on the Kagom\'e lattice, where
attention has so far been concentrated on the low-lying excitation
spectrum (see \cite{\WEBSLLP} and references therein).
\mn
In a final part, we have numerically computed the asymptotic
behaviour of the magnetization curve close to the transition to
saturation. The fundamental excitations associated to this
transition are hard-core bosons and one therefore finds a universal
behaviour: The characteristic DN-PT square-root in one dimension
\cite{\DzNe,\PoTa}, a linear behaviour with a logarithmic
correction in two dimensions \cite{\ZhNi,\Gluzman,\SSS} and a simple
linear behaviour in three and more dimensions \cite{\BaBra}.
This issue was studied for the transition to saturation because
the diagonalization simplifies considerably in this limit. In
particular, one can explicitly map the problem to a low-density
gas of hard-core bosons. In one dimension, where the nature of
the excitations is not really important, almost all second-order
transitions at plateau-boundaries are in the DN-PT universality
class. So, presumably the universality class observed at the transition
to saturation is also more general in higher dimensions. However,
there is also room for different behaviour since now the nature
of the fundamental excitations and the interactions are more
important.
\vskip 2 cm
\displayhead{Acknowledgments}
\mn
I am grateful to D.C.\ Cabra, M.\ Kaulke, P.\ Pujol, K.D.\ Schotte,
M.\ Troyer and M.E.\ Zhitomirsky for useful discussions.
Part of the numerical computations have been performed
on computers of the Max-Planck-Institut f\"ur Mathematik, Bonn-Beuel.
\vfill
\eject
\displayhead{References}
\mn
\bibitem{\Manou} E.\ Manousakis, {\it The Spin-${1 \over 2}$ Heisenberg
              Antiferromagnet on a Square Lattice and its Applications to the
              Cuprous Oxides}, Rev.\ Mod.\ Phys.\ {\bf 63} (1991) 1-62
\bibitem{\OYA} M.\ Oshikawa, M.\ Yamanaka, I.\ Affleck, {\it Magnetization
              Plateaus in Spin Chains: ``Haldane Gap'' for Half-Integer
              Spins}, Phys.\ Rev.\ Lett.\ {\bf 78} (1997) 1984-1987
\bibitem{\Totsuka} K.\ Totsuka, {\it Magnetization Processes in Bond-Alternating
              Quantum Spin Chains}, Phys.\ Lett.\ {\bf A228} (1997) 103-110
\bibitem{\CHP} D.C.\ Cabra, A.\ Honecker, P.\ Pujol, {\it Magnetization
              Curves of Antiferromagnetic Heisenberg Spin-${1 \over 2}$
              Ladders}, Phys.\ Rev.\ Lett.\ {\bf 79} (1997) 5126-5129
\bibitem{\CHPii} D.C.\ Cabra, A.\ Honecker, P.\ Pujol,
              {\it Magnetization Plateaux in $N$-Leg Spin Ladders},
              Phys.\ Rev.\ {\bf B58} (1998) 6241-6257
\bibitem{\Tot} K.\ Totsuka, {\it Novel Massive Ground States of Spin Chains in
              a Magnetic Field}, Eur.\ Phys.\ J.\ {\bf B5} (1998) 705-717
\bibitem{\CaGy} D.C.\ Cabra, M.D.\ Grynberg, {\it Ground State Magnetization of
              Polymerized Spin Chains}, Phys.\ Rev.\ {\bf B59} (1999) 119-122
\bibitem{\poly} A.\ Honecker, {\it A Strong-Coupling Approach to the
              Magnetization Process of Polymerized Quantum Spin Chains},
              preprint cond-mat/9808312, SISSA 93/98/EP,
              to appear in Phys.\ Rev.\ {\bf B}
\bibitem{\Korshunov} S.E.\ Korshunov, {\it Antiferromagnetic $XY$ Model on a
              Triangular Lattice: Ordered States in a Magnetic Field}, JETP
              Lett.\ {\bf 41} (1985) 641-643
\bibitem{\DoUi} Vik.S.\ Dotsenko, G.V.\ Uimin, {\it Phase Transitions in
              Two-Dimensional Antiferromagnetic $XY$ Models in External
              Fields}, J.\ Phys.\ C: Solid State Phys.\ {\bf 18} (1985)
              5019-5032
\bibitem{\Miyashita} S.\ Miyashita, {\it Magnetic Properties of Ising-Like
              Heisenberg Antiferromagnets on the Triangular Lattice}, J.\
              Phys.\ Soc.\ Jpn.\ {\bf 55} (1986) 3605-3617
\bibitem{\MiNi} H.\ Nishimori, S.\ Miyashita, {\it Magnetization Process in the
              Spin-$1/2$ Antiferromagnetic Ising-Like Heisenberg Model on the
              Triangular Lattice}, J.\ Phys.\ Soc.\ Jpn.\ {\bf 55} (1986)
              4448-4455
\bibitem{\ChuGo} A.V.\ Chubukov, D.I.\ Golosov, {\it Quantum Theory of an
              Antiferromagnet on a Triangular Lattice in a Magnetic Field},
              J.\ Phys.: Condensed Matter {\bf 3} (1991) 69-82
\bibitem{\BLLP} B.\ Bernu, P.\ Lecheminant, C.\ Lhuillier, L.\ Pierre, {\it
              Exact Spectra, Spin Susceptibilities, and Order Parameter of
              the Quantum Heisenberg Antiferromagnet on the Triangular
              Lattice}, Phys.\ Rev.\ {\bf B50} (1994) 10048-10062
\bibitem{\SuMa} N.\ Suzuki, F.\ Matsubara, {\it Phase Diagrams of the $S={1
              \over 2}$ Quantum Antiferromagnetic $XY$ Model on the
              Triangular Lattice in Magnetic Fields}, Phys.\ Rev.\ {\bf B55}
              (1997) 12331-12337
\bibitem{\LoNo} Y.E.\ Lozovik, O.I.\ Notych, {\it Magnetization Plateaux of
              Frustrated Antiferromagnet and Analogy with FQHE}, Solid State
              Communications {\bf 85} (1993) 873-877
\bibitem{\YaMue} M.S.\ Yang, K.-H.\ M\"utter, {\it The Two Dimensional
              Antiferromagnetic Heisenberg Model in the Presence of an
              External Field}, Z.\ Phys.\ {\bf B104} (1997) 117-123
\bibitem{\KoTa} M.\ Kohno, M.\ Takahashi, {\it Magnetization Process of the
              Spin-1/2 XXZ Models on Square and Cubic Lattices},
              Phys.\ Rev.\ {\bf B56} (1997) 3212-3217
\bibitem{\ZhNi} M.E.\ Zhitomirsky, T.\ Nikuni, {\it Magnetization Curve of a
              Square-Lattice Heisenberg Antiferromagnet}, Phys.\ Rev.\ {\bf
              B57} (1998) 5013-5016
\bibitem{\TrSa} M.\ Troyer, S.\ Sachdev, {\it Universal Critical Temperature
              for Kosterlitz-Thouless Transitions in Bilayer Quantum
              Magnets}, Phys.\ Rev.\ Lett.\ {\bf 81} (1998) 5418-5421
\bibitem{\WWB} F.Y.\ Wu, X.N.\ Wu, H.W.J.\ Bl\"ote, {\it Critical Frontier of
              the Antiferromagnetic Ising Model in a Magnetic Field: The
              Honeycomb Lattice}, Phys.\ Rev.\ Lett.\ {\bf 62} (1989)
              2773-2776
\bibitem{\IAG} T.\ Inami, Y.\ Ajiro, T.\ Goto, {\it Magnetization Process of
              the Triangular Lattice Antiferromagnets $RbFe(MoO_4)_2$ and
              $CsFe(SO_4)_2$}, J.\ Phys.\ Soc.\ Jpn.\ {\bf 65} (1996)
              2374-2376
\bibitem{\CoPe} M.F.\ Collins, O.A.\ Petrenko, {\it Triangular
              Antiferromagnets}, Can.\ J.\ Phys.\ {\bf 75} (1997) 605-655
\bibitem{\HTM} H.\ Nojiri, Y.\ Tokunaga, M.\ Motokawa, {\it Magnetic
              Phase Transition of Helical CsCuCl$_3$ in High Magnetic
              Field}, Journal de Physique {\bf 49}, Suppl.\ C8
              (1988) 1459-1460
\bibitem{\HKSprep} A.\ Honecker, M.\ Kaulke, K.D.\ Schotte, in preparation
\bibitem{\BoFi} J.C.\ Bonner, M.E.\ Fisher, {\it Linear Magnetic Chains with
              Anisotropic Coupling}, Phys Rev.\ {\bf 135} (1964) A640-658
\bibitem{\BoPa} J.B.\ Parkinson, J.C.\ Bonner, {\it Spin Chains in a Field:
              Crossover from Quantum to Classical Behavior}, Phys.\ Rev.\
              {\bf B32} (1985) 4703-4724
\bibitem{\Night} M.P.\ Nightingale, {\it Transfer Matrices, Phase Transitions,
              and Critical Phenomena: Numerical Methods and Applications},
              pp.\ 287-351 in: V.\ Privman (ed.),
              {\it Finite Size Scaling and Numerical Simulations of
              Statistical Physics}, World Scientific, Singapore (1990)
\bibitem{\BiLe} A.\ Bienenstock, J.\ Lewis, {\it Order-Disorder of
              Nonstoichiometric Binary Alloys and Ising Antiferromagnets in
              Magnetic Fields}, Physics Rev.\ {\bf 160} (1967) 393-403
\bibitem{\SWW} M.\ Schick, J.S.\ Walker, M.\ Wortis, {\it Phase Diagram of the
              Triangular Ising Model: Renormalizarion-Group Calculation with
              Application to Adsorbed Monolayers}, Phys.\ Rev.\ {\bf B16}
              (1977) 2205-2219
\bibitem{\Pert} A.\ Honecker, {\it A Perturbative Approach to Spectrum and
              Correlation Functions of the Chiral Potts Model}, J.\ Stat.\
              Phys.\ {\bf 82} (1996) 687-741
\bibitem{\WOH} Z.\ Weihong, J.\ Oitmaa, C.J.\ Hamer, {\it Square-Lattice
              Heisenberg Antiferromagnet at $T=0$}, Phys.\ Rev.\ {\bf B43}
              (1991) 8321-8330
\bibitem{\WOHii} J.\ Oitmaa, C.J.\ Hamer, Z.\ Weihong, {\it Quantum Magnets on
              the Honeycomb and Triangular Lattices at $T = 0$}, Phys.\ Rev.\
              {\bf B45} (1992) 9834-9841
\bibitem{\LeRu} P.W.\ Leung, K.J.\ Runge, {\it Spin-${1 \over 2}$ Quantum
              Antiferromagnets on the Triangular Lattice}, Phys.\ Rev.\ {\bf
              B47} (1993) 5861-5873
\bibitem{\Gluzman} S.\ Gluzman, {\it Two-Dimensional Quantum Antiferromagnet
              in a Strong Magnetic Field}, Z.\ Phys.\ {\bf B90} (1993) 313-318
\bibitem{\SSS} S.\ Sachdev, T.\ Senthil, R.\ Shankar, {\it Finite-Temperature
              Properties of Quantum Antiferromagnets in a Uniform Magnetic
              Field in One and Two Dimensions}, Phys.\ Rev.\ {\bf B50} (1994)
              258-272
\bibitem{\HoPa} R.P.\ Hodgson, J.B.\ Parkinson, {\it Bethe {\rm Ansatz} for
              Two-Deviation States in Quantum Spin Chains of Arbitrary $S$
              with Anisotropic Heisenberg Exchange}, J.\ Phys.\ C: Solid
              State Phys.\ {\bf 18} (1985) 6385-6395
\bibitem{\DzNe} G.I.\ Dzhaparidze, A.A.\ Nersesyan, {\it Magnetic-Field Phase
              Transition in a One-Dimen\-sional System of Electrons with
              Attraction}, JETP Lett.\ {\bf 27} (1978) 334-337
\bibitem{\PoTa} V.L.\ Pokrovsky, A.L.\ Talapov, {\it Ground State, Spectrum and
              Phase Diagram of Two-Dimensional Incommensurate Crystals},
              Phys.\ Rev.\ Lett.\ {\bf 42} (1979) 65-67
\bibitem{\Popov} V.N.\ Popov, {\it On the Theory of the Superfluidity of Two-
              and One-Dimensional Bose System}, Theor.\ Math.\ Phys.\ {\bf
              11} (1972) 565-573
\bibitem{\PopB} V.N.\ Popov, {\it Functional Integrals in Quantum Field
              Theory and Statistical Physics}, D.\ Reidel Publishing,
              Dordrecht, Holland (1983)
\bibitem{\BaBra} E.G.\ Batyev, L.S.\ Braginski\u{\i}, {\it Antiferromagnet in a
              Strong Magnetic Field: Analogy with Bose Gas}, Soviet Physics
              JETP {\bf 60} (1984) 781-786
\bibitem{\SGMK} M.\ Schmidt, C.\ Gerhardt, K.-H.\ M\"utter, M.\ Karbach, {\it
              The Frustrated Antiferromagnetic Heisenberg Model in the
              Presence of a Uniform Field}, J.\ Phys.: Condensed Matter {\bf
              8} (1996) 553-568
\bibitem{\OHA} K.\ Okunishi, Y, Hieida, Y.\ Akutsu, {\it Delta-Function Bose
              Gas Picture of $S=1$ Antiferromagnetic Quantum Spin Chains Near
              Critical Fields}, preprint cond-mat/9807266
\bibitem{\CHPzz} D.C.\ Cabra, A.\ Honecker, P.\ Pujol,
              {\it Magnetic Properties of Zig-Zag Ladders}, preprint
              cond-mat/9902112, BONN-TH-99-01, SISSA 10/99/EP, ETH-TH/99-01,
              ENSL-Th 02/99
\bibitem{\GJS} O.\ Golinelli, Th.\ Jolicoeur, E.S.\ S{\o}rensen, {\it
              Incommensurability in the Magnetic Excitations of the
              Bilinear-Biquadratic Spin-1 Chain}, preprint cond-mat/9812296,
              SPhT/98-120
\bibitem{\MBLW} G.\ Misguich, B.\ Bernu, C.\ Lhuillier, C.\ Waldtmann, {\it
              Spin Liquid in the Multiple-Spin Exchange Model on the
              Triangular Lattice: $^3$He on Graphite}, Phys.\ Rev.\ Lett.\
              {\bf 81} (1998) 1098-1101
\bibitem{\MoSaKu} T.\ Momoi, H.\ Sakamoto, K.\ Kubo, {\it Magnetization Plateau
              in a Two-Dimensional Multiple-Spin Exchange Model}, preprint
              cond-mat/9810118
\bibitem{\KuMo} K.\ Kubo, T.\ Momoi, {\it Ground State of a Spin System with
              Two- and Four-Spin Exchange Interactions on the Triangular
              Lattice}, Z.\ Phys.\ {\bf B103} (1997) 485-489
\bibitem{\GrSh} B.\ Gr\"unbaum, G.C.\ Shephard, {\it Tilings and Patterns},
              W.H.\ Freeman and Company, New York (1987)
\bibitem{\WEBSLLP} Ch.\ Waldtmann, H.-U.\ Everts, B.\ Bernu, P.\ Sindzingre,
              C.\ Lhuillier, P.\ Lecheminant, L.\ Pierre, {\it First
              Excitations of the Spin 1/2 Heisenberg Antiferromagnet on the
              Kagom\'e Lattice}, Eur.\ Phys.\ J.\ {\bf B2} (1998) 501-507
\vfill
\end